\PassOptionsToPackage{table,xcdraw}{xcolor}
\documentclass[conference]{IEEEtran}
\usepackage{svg}
\usepackage{booktabs}
\usepackage{multirow}
\usepackage[noline, ruled,linesnumbered]{algorithm2e}
\usepackage{url}
\usepackage{hyperref}
\usepackage{amsmath}
\usepackage{amsfonts}
\usepackage{cite}
\let\labelindent\relax  % Undefine \labelindent
\usepackage{enumitem}
\usepackage{fancyhdr}

\ifCLASSINFOpdf
\else
\fi
\definecolor{ndssblue}{RGB}{0,61,165} % Adjust RGB values for NDSS blue
\fancypagestyle{firstpage}{
    \fancyhf{} % Clear all header and footer fields
    \fancyhead[C]{\textcolor{ndssblue}{\Large \textbf{This paper has been accepted to NDSS 2025}}} % Center-aligned header with larger deep blue font
}

\begin{document}


\title{Revisiting Concept Drift in Windows  \\ Malware Detection: Adaptation to Real \\Drifted Malware with Minimal Samples}

% \author{\IEEEauthorblockN{Adrian Shuai Li}
% \IEEEauthorblockA{Purdue University\\
% li3944@purdue.edu}
% \and
% \IEEEauthorblockN{Arun Iyengar}
% \IEEEauthorblockA{Cisco Research\\
% ariyenga@cisco.com}
% \and
% \IEEEauthorblockN{Ashish Kundu}
% \IEEEauthorblockA{Cisco Research\\
% ashkundu@cisco.com}
% \and
% \IEEEauthorblockN{Elisa Bertino }
% \IEEEauthorblockA{Purdue University\\
% bertino@purdue.edu}}


%for over three affiliations, or if they all won't fit within the width of the page, use this alternative format:
\author{\IEEEauthorblockN{Adrian Shuai Li\IEEEauthorrefmark{1},
Arun Iyengar\IEEEauthorrefmark{2},
Ashish Kundu\IEEEauthorrefmark{3}, 
Elisa Bertino\IEEEauthorrefmark{1}}
\IEEEauthorrefmark{1}Purdue University, \IEEEauthorrefmark{2}Intelligent Data Management and Analytics, LLC, \IEEEauthorrefmark{3}Cisco Research \\
\IEEEauthorrefmark{1}\{li3944, bertino\}@purdue.edu, \IEEEauthorrefmark{2}aki@akiyengar.com, \IEEEauthorrefmark{3}ashkundu@cisco.com}

% \IEEEoverridecommandlockouts
% \makeatletter\def\@IEEEpubidpullup{6.5\baselineskip}\makeatother
% \IEEEpubid{\parbox{\columnwidth}{
% 		Network and Distributed System Security (NDSS) Symposium 2025\\
% 		24-28 February 2025, San Diego, CA, USA\\
% 		ISBN 979-8-9894372-8-3\\
% 		https://dx.doi.org/10.14722/ndss.2025.240830\\
% 		www.ndss-symposium.org
% }
% \hspace{\columnsep}\makebox[\columnwidth]{}}

\maketitle

\thispagestyle{firstpage}

\begin{abstract}
In applying deep learning for malware classification, it is crucial to account for the prevalence of malware evolution, which can cause trained classifiers to fail on drifted malware. Existing solutions to address concept drift use active learning. They select new samples for analysts to label and then retrain the classifier with the new labels. Our key finding is that the current retraining techniques do not achieve optimal results. These techniques overlook that updating the model with scarce drifted samples requires learning features that remain consistent across pre-drift and post-drift data. The model should thus be able to disregard specific features that, while beneficial for the classification of pre-drift data, are absent in post-drift data, thereby preventing prediction degradation. In this paper, we propose a new technique for detecting and classifying drifted malware that learns drift-invariant features in malware control flow graphs by leveraging graph neural networks with adversarial domain adaptation.  We compare it with existing model retraining methods in active learning-based malware detection systems and other domain adaptation techniques from the vision domain. Our approach significantly improves drifted malware detection on publicly available benchmarks and real-world malware databases reported daily by security companies in 2024.  We also tested our approach in predicting multiple malware families drifted over time. A thorough evaluation shows that our approach outperforms the state-of-the-art approaches.

\end{abstract}


% \fi
%%\IEEEpeerreviewmaketitle

\section{Introduction}\label{intro}

%Recently, we have witnessed a surge of research efforts that leverage deep learning to automate malware 
%classification task, 

Recent automatic malware classification methods use deep learning (DL) techniques with various feature types, including static features~\cite{ahmadi2016novel, aghakhani2020malware, nataraj2011malware, vasan2020imcfn, singh2019malware, bhodia2019transfer,yan2019classifying, someya2023fcgat,xu2019droidevolver, mariconti2016mamadroid} and dynamic features~\cite{yin2023method, hou2023proteus}, and their combinations~\cite{dambra2023decoding}. 
DL has shown noticeably better performance than the traditional signature-based approach and machine learning (ML) methods. However, a major challenge in the use of DL is the concept drift,  
%The use of deep learning techniques for classifying malware faces a challenge known as concept drift, 
wherein the distribution of test data diverges from the one of the original training data. While this issue is not unique to malware—it is akin to the ``out-of-distribution'' problem in DL—the intricacies of the malware evolution make domain drift particularly complex and crucial. %Malware classifiers are deployed in dynamic, hostile environments. 
Malware continually evolves, adopting new paradigms to evade detection and maximize the damage. %New variants are created as new exploits discovered.
Attackers further complicate detection by creating adversarial samples through code mutations and injection of dummy code. With the development of generative AI, research has shown that the attack can even jailbreak large language models for malicious code generation~\cite{pa2023attacker,xie2023defending}. 

Traditional methods to handle concept drift in malware classification involve labeling new samples and retraining the model. However, this process is time-consuming, costly, and sometimes impractical due to analysts’ limited capacity for daily sample labeling. Another approach involves using pseudo-labels to provide noisy labels for drifted malware, which is then used to update the model~\cite{kan2021investigating, xu2019droidevolver}. However, this method relies heavily on the accuracy of the initial estimate and is prone to negative feedback loops if the process is not designed properly. The state-of-the-art solutions use active learning to adapt to concept drift. They deliberately select important labels crucial for learning new malware distribution and then retrain the classifier with those new labels. There are many schemes for selecting which samples to label ~\cite{pendlebury2019tesseract, narayanan2016adaptive, narayanan2017context,barbero2022transcending, yang2021cade, chen2023continuous}, with the goal of reducing the amount of manual labeling effort needed to achieve a good performance. 

Past work has made progress in the detection of drifted malware.  Nonetheless, past approaches have predominantly used basic retraining techniques for model 
updating. Chen et al.~\cite{chen2023continuous} were the first to distinguish between two prevalent model updating strategies in active learning. The first strategy, known as cold-start learning, involves training a fresh model each time new labels are introduced. The second strategy, referred to as warm-start learning, continues training an existing model with new samples. However, our experimental findings indicate that neither strategy yields optimal performance when only a few new samples are available. 

We envision that addressing concept drift, particularly with scarce labeled samples,  requires learning features that remain consistent before and after the drift. Malware detection models should thus be able to disregard features from the pre-drift data that may not be present in the post-drift data. Neural networks are prone to ``cheating'' where they resort to shortcuts during prediction~\cite{mit}, leading to prediction failures when tested with data devoid of such shortcut information~\cite{li2024transfer}. This problem extends to malware prediction systems as well. %Depending on the type of malware, assembly instructions often exhibit intricate characteristics.
For instance, some malware samples may primarily consist of a few define directives for reserving variable storage space due to packing and obfuscation. These samples predominantly contain instructions like db, dw, and dd, which allocate byte, word, and double word, respectively. If a neural network is trained with many of these malware samples, it might leverage this pattern to predict an input as malicious upon encountering a series of these instructions. Consequently, it might incorrectly classify a malware sample with fewer of these operation codes but more API calls as benign. Likewise, a benign software that employs packing tools could be misclassified as malicious, resulting in a false positive
due to the prevalence of define directive instructions in its code. Therefore, an adaptive model should avoid using domain-specific features and instead acquire knowledge of common characteristics in malware executables. Both the cold-start and warm-start learning techniques do not exhibit these qualities.

\textbf{Our Method.} In this paper, we introduce a novel graph-based adversarial domain adaptation (DA) method to %combat 
address malware drift, coupled with a new method for graph-based clustering that identifies statistically distant malware clusters for evaluation. At a high level, warm-start learning enhances cold-start learning by using a model that has been previously trained on existing malware. Our approach learns a new model directly from the existing and drifted malware samples simultaneously.
To facilitate effective knowledge transfer to new samples, our methodology focuses on learning an intermediate representation containing information that remains consistent before and after the drift while still being sufficient to make a good classification.  %We leverages the control flow and code semantics extracted from malware executables.  
%First we use a two-pass method to extract CFGs from malware code, capturing its semantic and structural features~\cite{yan2019classifying}.
We use Control Flow Graphs (CFGs) derived from malware assembly, as they are more comprehensive and contain significantly more nodes than function call graphs. Then, we use a pre-trained assembly model~\cite{li2021palmtree} to generate embeddings for instructions in CFGs for neural network training. %PalmTree is trained on large-scare binary corpora, enabling it to capture the semantics of assembly language. 
%In addition, as we would like to make the most use of graph information, including structures, node attributes and labels, to learn effective graph representations, we introduce a training task focused on learning a representation containing the aforementioned information through the prediction of graph labels. This training task, referred to as Label Prediction, is based on the Graph Isomorphism Network (GIN)~\cite{xu2018powerful}, widely recognized as one of the most expressive graph neural networks available.
We introduce a training task, \textit{Label Prediction (LP)}, based on the Graph Isomorphism Network (GIN)~\cite{xu2018powerful}, to learn graph representations from graph structures, code semantics, and labels. To address the shift between old and new malware samples, we introduce another learning task involving the training of two networks through minimax optimization to predict the input domain (pre-drift or post-drift). This training task referred to as \textit{Adversarial Training (AT)}, draws inspiration from generative adversarial networks~\cite{goodfellow2020generative} and DA for images~\cite{li2024transfer, ganin2016domain}.  Our model automatically learns domain-invariant representations through those two training tasks using CFGs as the input. The domain-invariant representations are indistinguishable regarding whether they originate from pre-drift or post-drift data.  If this representation allows us to achieve strong classification performance on the pre-drift data,  it will also improve generalization on the post-drift data. 

We also emphasize the importance of proper evaluation of techniques tackling concept drift. Previous research has frequently overlooked the validation of actual drift occurrence (particularly on research datasets), potentially resulting in an overestimation of the algorithm's predictive accuracy~\cite{kumar2021mcft, ma2021comprehensive, vasan2020imcfn, narayanan2016adaptive, narayanan2017context, xu2019droidevolver,wang2022efficient}. Despite varying experiment designs, these studies rely on the labels of malware families and typically exclude one label to serve as the ground truth for the “unseen family”. However, closely related malware families might enable the algorithm to predict one family exceptionally well, even when trained on another, an issue similar to the temporal bias issue studied in~\cite{pendlebury2019tesseract}. In one of the 
most used Big-15 benchmarks~\cite{ronen2018microsoft}, the distance between malware samples from different families is strikingly small. In this paper,  we advocate for assessing the extent to which malware characteristics have changed within the dataset. We do so in all our experiments and provide distinct cluster labels for datasets where the original labels are not well separated, along with a new graph-based clustering algorithm for generating these clusters. For the clustering algorithm, we use an ensemble of clustering predictors. The clusters are then generated through a weighted consensus process that takes into account the performance of each predictor. Several clustering performance indexes demonstrate that this approach improves cluster assignments, effectively separating distant samples and grouping close ones.

%\vspace{-2mm}

\textbf{Evaluation.} We conduct experiments to evaluate our approach and compare it with leading graph-based malware classification models in both cold-start and warm-start learning settings. We also explore other DA methods to determine which DA techniques are most efficient for malware detection. Our initial experiments are conducted on malware classification research benchmarks, with one family as the target and the rest as pre-drift data. We used original family labels in Big-15~\cite{ronen2018microsoft} and then with our cluster labels. We find that the model trained with warm-start learning has a performance decline of at most $5.8\%$ - the highest-upon evaluation with cluster labels of Big-15. Conversely, our method experiences only a minor $0.5\%$ performance decline on this dataset and outperforms all the others. 
We evaluate our design choice of using CFGs for malware representation against content-based~\cite{aghakhani2020malware, dambra2023decoding}  and 
image-based~\cite{nataraj2011malware, vasan2020imcfn, singh2019malware, bhodia2019transfer} representations on Big-15 research benchmark.  Our adaptation technique proves effective across all representations, and the best results are achieved when combined with our graph representation. We also find that in both cold-start and warm-start learning scenarios, overall the graph representation surpasses the other two representations. 

We use the MB-24 malware dataset, which we collected from the MalwareBazaar daily feed between March and September 2024, for real-world malware evaluation. Data from March to May serves as pre-drift data. In July, we labeled a fixed set of samples for post-drift training, updating the model. This updated model is tested using August data. Then, we labeled a few samples in August for further updates and tested the new model with September data. On average, we matched the accuracy of the upper bound results obtained through supervised training for each testing month while reducing the labeling effort by $5$X. In subsequent experiments using the MalwareDrift dataset, our method improved family-level classification by $9-14\%$ over the state-of-the-art with just $10$ new samples per family. This was achieved in a closed-set DA scenario, where pre-drift and post-drift samples are from the same malware families. We further demonstrate that our approach effectively extends to open-set DA, where new malware families appear in the post-drift data.  Our final experiment demonstrates that our DA method performs better by learning features that reduce the distribution divergence between pre-drift and post-drift data.
%\vspace{-2mm}

\textbf{Contributions.}
%\vspace{-2mm}
This paper has three main contributions.
\begin{itemize}
    \item We introduce a novel model training method based on CFG and DA to address the concept drift problem in Windows malware classification. Our method fits in both closed and open-set scenarios.
    \item We highlight the limitations of previous work that conducted evaluations on research datasets without verifying actual drift occurrences. We propose a new graph-based clustering method that computes statistically distinct malware clusters to eliminate bias.
    \item We extensively evaluate many previously proposed methods for model updating to determine the most effective solution—encompassing both malware representation and training methods—against malware drift, an area that has not been studied before. The results demonstrate significant improvements over previous work. We have released the code and data to support future research, with details provided in the Artifact Appendix.
\end{itemize}

% We release the code at \url{https://anonymous.4open.science/r/ccs_2024-8B1E/README.md} to support future research.

\section{Methodology}\label{method}

% In this work, we investigate domain adaptation (DA) for control flow graphs (CFG) of malware and benign software binaries. 

We leverage the information from existing labeled malware samples (source data\footnote{The terms ``source domain" and ``source data" are used interchangeably.}) to aid in the classification of partially labeled new malware samples (target data\footnote{The terms ``target domain" and ``target data" are used interchangeably.}).  %In practical scenarios, the creation of source and target data can be accomplished by utilizing the drift detection techniques referred to in Section~\ref{related_work},  which is beyond the scope of this work. 
In this section, we first formally define the research problem and introduce the notations used throughout the rest of the paper. Next, we demonstrate how our suggested approach effectively addresses the challenges outlined in Section~\ref{intro}.
%This will be presented with an overview followed by an in-depth exploration of the system design.

\subsection{Problem definition}

\begin{figure*}[t]
  \centering
  \includegraphics[width=0.9\linewidth]{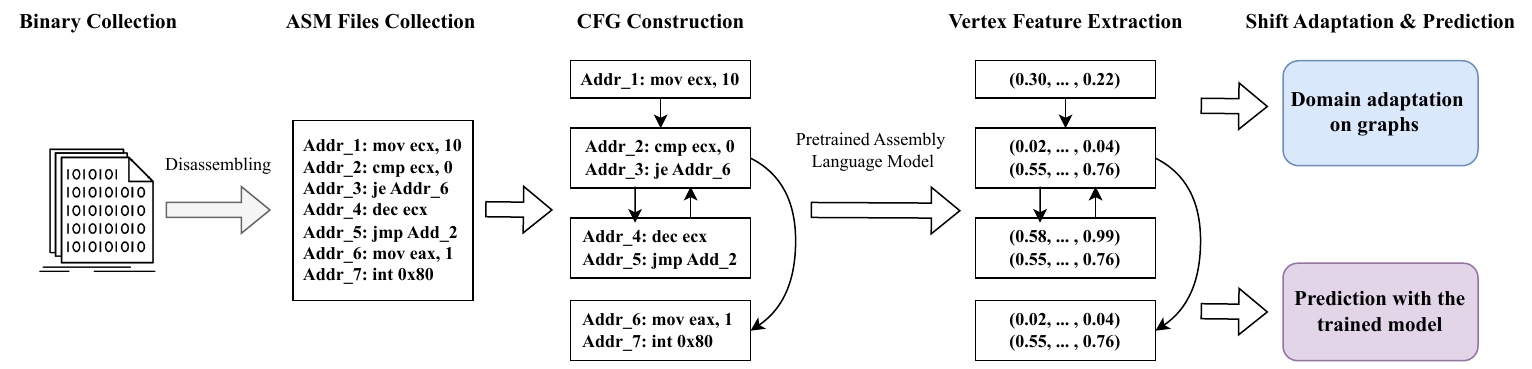}
  \caption{%System design. 
 Overview of our approach: we show the assembly code on the left and the corresponding control flow graph on the right. }
  \label{data}
\end{figure*}

Given a binary, we use its representation as a control flow graph (CFG).  A CFG is a directed graph where vertices represent sequences of assembly instructions, and edges represent the execution flow. Each vertex represents a basic block; in what follows, we use the terms ``basic block" and ``node"
%, known as a basic block or a node, is used 
interchangeably.
%within this paper. 
Figure~\ref{data} shows a binary code snippet alongside its corresponding CFG. Each instruction is converted to a vector. We average the instruction embeddings to obtain the node attribute.
%\ElisaText{You need to mention that each node has attributes}

The source data is represented as  $G^s = \{G^{s}_{i}\} = \{(X^s_i, A^s_i, Y^s_i)\}$, where $X^s_{i} \in \mathbb{R}^{n^s_{i} \times m^s}$ is the node attribute matrix for $G^{s}_{i}$ with $n^s_i$ being the number of nodes and $m^s$ being the number of node attributes in the source data. Additionally, ${A^s_i} \in \mathbb{R}^{n^s_i \times n^s_i}$ is the adjacency matrix with $A^s_{i}(p,q)$ 
%COMMENT-FINAL: I think that the notation $A^s_{pq}$ is not correct. If by this notation you want to specify which is the content of the element of the adjaceny matrix, you should say something like
%"$A^s_{i}(p, q)$
denoting the number of edges between node $p$ and $q$, and ${Y^s_i}$ is the one-hot encoding of the classification label for $G^{s}_{i}$. 

Similarly, the target data is represented as $G^t = \{G^{t}_{i}\} = \{({X^t_i}, {A^t_i}, {Y^t_i})\}$, where ${X^t_i} \in \mathbb{R}^{n^t_i \times m^t}$ is the the node attribute matrix for $G^{t}_{i}$ with $n^t_i$ being the number of nodes and $m^t$ being the number of node attributes,  ${A^t_i} \in \mathbb{R}^{n^t_i \times n^t_i}$ is the adjacency matrix, and ${Y^t_i}$ is the one-hot encoding of the label for $G^{t}_{i}$. Furthermore, let $|G^{s}_{i}|$ represent the number of samples from the source domain and $|G^{t}_{i}|$ represent the sample size in the target domain. We assume that $|G^{s}_{i}| \gg |G^{t}_{i}|$. 

The source and target data contain the same attributes, that is $m^{s} = m^{t}$.  The number of attributes is adjustable and determined in Section~\ref{vertex}, where one can indicate the dimension of the node 
attribute matrix.
%\ElisaText{here do you mean the "node attribute matrix"? If so, please add "matrix"}
% The number of attributes is modifiable and set during the data preprocessing phase, allowing the specification of feature vector dimensions for the node attribute matrix.

We now formally state our research problem as follows: \textit{ A divergence exists between the source and target malware data, yet the label space remains consistent ($Y^s = Y^t$). Our main objective is to develop a classifier that can effectively identify drifted malware in the target domain with very few labeled graphs from the target. }  %To accomplish this, we make use of the graph structure and node attributes of CFGs from exisiting malware binaries in the source domain, which offers a significant amount of labeled data to aid in the learning process.
%\ElisaText{It seems that in the above problem we do not mention that the target has very few labeled data.}
We formulate our research problem as a closed-set prediction task where the source and target have the same label space. In this work, we also show that our approach can be extended to fit in the open-set scenario where the target may have new malware classes not in the source (see Section~\ref{openset} for details).

\begin{figure}[t]
  \centering
  \includegraphics[width=\linewidth]{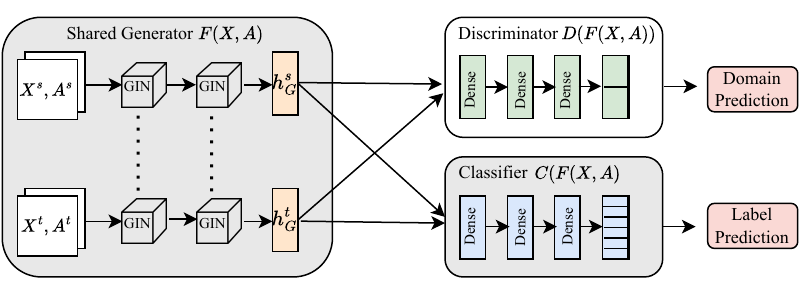}
  \caption{The inputs are source and target graph data, represented as node attribute matrix ($X^{s/t}$) and adjacency matrix ($A^{s/t}$). We obtain those two matrices for each graph after the Vertex Feature Extraction step in Figure~\ref{data}. The training process is modeled as a minimax game between the generator and the discriminator. $h^s_G$ and $h^t_G$ denote the graph-level representations corresponding to the source and target inputs, respectively. Following the training process, the discriminator fails to discern the domain distinction solely based on $h^s_G$ and $h^t_G$. At the same time, they retain useful information crucial for achieving good classification in both domains. }
  \label{arch}
\end{figure}

\subsection{Overview}

Figure~\ref{data} illustrates the design of our approach, which is composed of three key components: CFG Construction from ASM Files, Vertex Feature Extraction, and Shift Adaptation. Initially, we disassemble the malware binary to extract the CFG from the assembly code.   Each node in the graph represents a basic block in the assembly code, while edges represent jumps in the control flow. Moving to the second component, we employ a pre-trained language model to generate high-quality node embeddings.  After that, we train our model with source data and limited target data (see Figure~\ref{arch}). The model comprises two main components: the domain prediction component, consisting of a generator and a discriminator, and the classification component, composed of a generator and a classifier. During the optimization process, the shared generator can learn features that combine class 
distinctiveness and domain invariance. This empowers the classifier to classify data in the target domain with the assistance of the source data and only a limited set of labeled target data. Once the model has been trained, we deploy it for prediction.
%In Section~\ref{cfg}, 
In what follows, we introduce the approach for constructing CFGs, followed by the vertex feature extraction process. Finally, we present our method for training a graph neural network (GNN) with adversarial DA for drift malware adaptation.
%in Section~\ref{design}.

\subsection{CFG construction from disassembly}\label{cfg}
We first disassemble binary files using IDA Pro. We use the open-source code from MAGIC~\cite{yan2019classifying} for CFG extraction.  The algorithm employs a two-pass traversal methodology. %, which reads a .asm file as input and outputs the CFG.
Initially, the file is processed to create a mapping from addresses to instructions.  Instructions are associated with four tags, i.e., \{\verb"start",
\verb"branchTo", \verb"fallThrough", \verb"return"\},  which are used by the second pass. % for creating code blocks.
In the first pass, each instruction is visited and its associated tags are updated accordingly. The second pass is dedicated to creating basic blocks and edges between these blocks based on the assigned tags.

\subsection{Vertex feature extraction}\label{vertex}

To apply deep learning to the CFGs, the first step is to extract feature vectors for the instructions, as GNNs cannot directly operate on raw instructions. Among the available approaches – directly feeding raw bytes, employing manually designed features, or automatically generating vector representations for individual instructions using a designated representation model – 
the preference is given to the third approach.  This approach generally yields better-quality embeddings without requiring a manual selection of the features. PalmTree~\cite{li2021palmtree} is a pre-trained assembly language model based on BERT for general-purpose instruction representation learning. Experimental results~\cite{li2021palmtree} show its efficacy in generating high-quality instruction embeddings for various downstream binary analysis tasks. %such as binary similarity detection, function type signatures analysis, and value set analysis. %PalmTree uses three training tasks, each works at different granularity levels to effectively capture internal formats, contexual control flow dependency, and data flow dependency of instructions~\cite{li2021palmtree}.   Note that one can use other techniques  and models to generate instruction representations.
Hence, we use the PalmTree model with pre-trained parameters to generate instruction embeddings.

\subsubsection{Normalization} In NLP, the Out-of-Vocabulary (OOV) problem occurs when a token encountered during inference is absent in the vocabulary the model was trained on.  The OOV issue in assembly code encoding is particularly challenging given the diverse nature of codebases and the potential for encountering unseen strings and constant numbers. To address this issue, our initial step involves normalizing the CFG: strings are substituted with a designated token \verb"[str]".  Constants, which contain a minimum of five hexadecimal digits, are replaced by a specific token, \verb"[addr]". Smaller constants remain intact and are encoded as one-hot vectors. These strategies align with those utilized during PalmTree's training.

\subsubsection{Node feature embedding} After normalization, we apply the PalmTree model to generate an embedding for each instruction. We still need to aggregate vectors to obtain a single feature vector to be associated with a vertex.  We adopt an efficient yet effective solution based on an analysis of the related literature~\cite{massarelli2019investigating}. We aggregate all the instruction embeddings and compute an unweighted mean vector. This process is repeated for all basic blocks, yielding the final CFG. Each processed CFG  is represented by $ G = (X, A, Y)$, where $X\in \mathbb{R}^{n \times m}$ is the node attribute matrix for $G$ with $n$ being the number of nodes and $m$ being the dimension of each node feature vector,  $A \in \mathbb{R}^{n \times n_i}$ is the adjacency matrix, and $Y$ is the one-hot encoding of the label for $G$.

\vspace{-1mm}
\subsection{Shift adaptation: model design}\label{design}

\subsubsection{Representation learning}
Learning with graphs requires effectively representing their structures and node attributes. % We use GIN since it has been demonstrated representationally more powerful on many graph classification benchmarks ~\cite{xu2018powerful}. 
GNNs are an effective framework for the representation learning of graphs. They generally adopt a recursive aggregation approach where each node combines the feature vectors of its neighbors to derive its updated feature vector. With $k$ iterations of aggregation, a node is represented by a node feature vector, encapsulating the structural information within its k-hop neighborhood. The graph-level representation is achieved via a pooling function. Formally,  the k-th iteration of a GNN is constructed as:  
\begin{align}
  h_v^{(k)} &= \text{COMBINE}\left(h_v^{(k-1)}, a_v^{(k)}\right) \\
  a_v^{(k)} &= \text{AGGREGATE}\left(\left\{ h_u^{(k-1)}, \forall u \in \mathcal{N}(v) \right\}\right)
\end{align}
where $h_v^{(k)}$ is the feature vector of node $v$ at the $k$-th iteration.  We initialize $h_v^{(k)} = x_v$, where $x_v$ is the feature vector for node $v$.  $a_v^{(k)}$ denotes the aggregation of features vectors from $v$'s neighbours at the $(k-1)$-th iteration. $\mathcal{N}(v)$   are the neighbours to $v$. The selection of  $\text{AGGREGATE}$ and $\text{COMBINE}$ differs among GNN variants.  In Graph Isomorphism Network (GIN), which we use in this work, these steps are integrated as follows:
\begin{align}
h_v^{(k)} = \text{MLP}\left(\left(1+ \epsilon ^{(k)}\right) h_{v}^{(k-1)} + \sum_{u \in N(v)} h_{u}^{(k-1)} \right)
\end{align}
where $\epsilon$ represents a learnable parameter. We designate $\epsilon$ as 0, a configuration referred to as GIN-0 in~\cite{xu2018powerful}.

The graph representation $h_G$ is obtained by aggregating node features from the final iteration $K$:
\begin{align}
h_G = \text{READOUT}\left( \{h_v^{(K)} | v \in G\}\right)
\end{align}
$\text{READOUT}$ can be either a summation function or a graph-level pooling function. It is important to note that relying solely on node representations from the final layer may limit performance, as features from earlier iterations can sometimes generalize better~\cite{xu2018powerful}. To address this, we adopt a method similar to Jumping Knowledge Networks, wherein we concatenate graph representations across all iterations:
\begin{align}
h_G = \text{CONCAT}\left(\text{READOUT}\left( \{h_v^{(k)} | v \in G\}\right) | k = 0, 1, ...,K\right)
\end{align}
In the next subsection, we present the loss function designed to obtain graph representation $h_G$.
\subsubsection {Loss functions}

 A main goal for the successful DA is to generate a domain-independent representation of the data from different domains, that is, graph representations ($h_G$) from the source domain that are similar to those from the target domain.  This allows classifiers trained on graphs from the source domain to generalize to the target domain, as the inputs to the classifier are invariant with respect to the domain of origin. To generate such representation,  we leverage the domain adversarial training approach~\cite{li2023maximal, ganin2016domain} that adversarially trains two neural networks to ensure the representation is domain invariant. Those two networks serve as a discriminator and a generator, respectively. The generator is trained adversarially to maximize the discriminator's loss. To learn distinctive graph representations for classification, the generator is also trained with a label predictor.

More specifically,  the source inputs are given by $\{({X^s_i}, {A^s_i}, {Y^{s}_i}, d^s_i)\}$, where each element represents node matrix, node adjacency matrix, class label, and domain label.  The target is given by $\{({X^t_i}, {A^t_i}, {Y^{t}_i}, d^t_i)\}$. The domain label $d_i \in \{0,1\}$ is the ground truth domain label for sample $i$. Let $F(X,A;\theta_{f})$ be the generator parameterized by $\theta_{f}$,  which maps a node adjacency matrix $A$ and a node attribute matrix $X$ to a hidden graph representation,  $h_{G}$, representing graph features that are common across domains. Let $D(h_G;\theta_{d})$ be the discriminator that maps a hidden representation  $h_{G}$ to the domain-specific prediction. Finally, $C(h_G;\theta_{c})$  represents a classifier, parameterized by $\theta_{c}$ that maps from a hidden representation $h_G$ to the task-specific prediction. The resulting model is shown in Figure~\ref{arch}.

 % Source inputs is given by $\{({X^s_i}, {A^s_i}, {Y^{s}_i}, d^s_i)\}$ where each element represents node matrix, node adjacency matrix, class label and the domain label. Similarly, the target is given by $\{({X^t_i}, {A^t_i}, {Y^{t}_i}, d^t_i)\}$. The domain label is $0$ for the source domain and $1$ for the target domain. 

Inference in our model is given by $\hat{Y} = C(F(X ,A ; \theta_f); \theta_c)$ and $\hat{d} = D(F(X ,A ; \theta_f); \theta_d)$  where $\hat{Y}$ is the label prediction and $\hat{d}$ is the domain prediction. The goal of training is to minimize the following loss with respect to parameters $\theta_f, \theta_c, \theta_d$: 
\begin{align}
min_{\theta_f, \theta_c} &\{\gamma \mathcal{L}_g + \mathcal{L}_c \}\label{loss_f} \\
min_{\theta_d} &\{\mathcal{L}_d \}
\label{eq7}
\end{align}
where $\gamma$ is the weight that controls the $\mathcal{L}_g$ term. %The goal of training the discriminator is to minimize the following loss with respect to parameters $\theta_d$. 
The classification loss $\mathcal{L}_c$ trains the model to predict the output labels. Because we assume that the target domain has limited labels, the loss applies to both domains, and it is defined as follows:
\begin{align}
\mathcal{L}_c=  -  \sum_{i=1}^{N_s} {Y}_i^s \cdot log \hat{Y}_i^s -  \lambda \sum_{i=1}^{N_t} {Y}_i^t \cdot log \hat{Y}_i^t
\end{align}
where $N_s$ represents the number of samples from the source domain, $\hat{Y}_i^s$ is the softmax output of the model: $\hat{Y}_i^s = C(F(X^s_i ,A^s_i))$. $N_t$ represents the number of samples from the target domain, $\hat{Y}_i^t$ is the softmax output of the model: $\hat{Y}_i^t = C(F(X^t_i ,A^t_i))$. We use $\lambda$ as the penalty coefficient for the loss value obtained from target data.

The discriminator loss trains the discriminator to predict whether the output of $F$ is generated from the source or the target domain.   Let $\hat{d}_i^s = D(F(X^s_i ,A^s_i))$ and $\hat{d}_i^t = D(F(X^t_i ,A^t_i))$ be the domain predictions from samples of source and target, respectively. The discriminator loss is also applied to both domains, and it is defined as follows: 
\begin{equation}
\mathcal{L}_d= 
 -  \sum_{i=1}^{N_s + N_t} \left[ {d}_ilog \hat{d}_i + (1-{d}_i)log (1-\hat{{d}}_i) \right] 
\end{equation}%
Finally, the generator loss encourages the hidden graph representations $h_G^s$ and $h_G^t$ from the generator to be as similar as possible so the discriminator cannot predict the domain of the representation. This is achieved via adversarially training the generator so that parameters  $\theta_f$ are optimized to reduce the domain classification accuracy.  Essentially, we minimize the loss for the domain prediction task with respect to $\theta_d$, while maximizing it with respect to $\theta_f$. Hence, the generator loss is defined with inverted ground truth domain labels: 
\begin{equation}
\mathcal{L}_g= 
 -  \sum_{i=1}^{N_s + N_t} \left[ (1-{d}_i)log \hat{d}_i + d_ilog (1-\hat{{d}}_i) \right]
\end{equation}%
\subsubsection{Model training}
Training $F$ consists of optimizing $\mathcal{L}_g$ and $\mathcal{L}_c$, since we want to minimize the domain classification accuracy and maximize label classification accuracy. The discriminator is trained with $\mathcal{L}_d$ to maximize domain classification accuracy. The classifier is trained with $\mathcal{L}_c$ to maximize label prediction accuracy. The training algorithm follows the mini-batch gradient descent procedure.  More specifically,  the following steps are executed after creating the mini-batches. The generator updates its weight to minimize $\gamma \mathcal{L}_g + \mathcal{L}_c$. The classifier updates its weight to minimize classification loss. The discriminator weights remain frozen during this step. Then, the discriminator updates its weight to minimize discriminator loss. Upon the model’s convergence, we can achieve graph representations that are both discriminative of the class and invariant to the domain. To classify graphs in the target, one can obtain prediction by running  $C(F(X,A))$.

\section{Generating drifted malware clusters}\label{gca}
%\ElisaText{I have slighly changed this initial sentence. See if you like.}
Past approaches to malware drift were evaluated 
%Previous malware drift analyses evaluated their approaches 
%by designing experiments 
using existing malware research datasets. Despite varying experiment designs, they all rely on the original malware family labels 
%of malware 
for their analysis. In this paper, we do so as well. However, we discovered that despite being labeled as different families, malware samples exhibit highly similar characteristics in one of the most used benchmarks. This phenomenon is evidenced by the data in Table~\ref{metric}, which shows that the distances between malware samples from different families are remarkably small. Evaluation relying on those family labels is likely to overestimate the accuracy of the prediction. Consequently, we developed and implemented a graph-based clustering algorithm to assign cluster labels to malware, thereby increasing inter-cluster distance and amplifying domain shift. We conducted our experimentation on Big-15 in two scenarios to demonstrate the effect of performance overestimation: one evaluation using the original labels (Section~\ref{exp_origin}), and another using the labels obtained by using clustering (Section~\ref{exp_cluster}).

% \textbf{Evaluation based on original labels.} The evaluation was conducted using the original labels of the dataset. We employed the ``leave-one-out" approach, where we designated one family label as the target domain and treated the remaining family labels as the source domain.

% \textbf{Evaluation based on cluster labels.} We assign new labels to the samples using the graph clustering approach detailed below. Then we apply the same ``leave-one-out" method. This task is more difficult due to the significantly increased distances between clusters compared to those with the initial labels, as shown in Table~\ref{metric}.

%COMMENT-FINALL as Section III has only one subsetion, then we should remove the subsection title. Usually you split a section into subsections, only when there are at least two subsections. As Section III has only one (titled Graph-based clustering algorithm), we would remove the subsection title below. The subsubsections will then become subsections. 
%\subsection{Graph-based clustering algorithm}\label{gca}
The graph-based clustering algorithm comprises two primary components: (1) the graph embedding component, responsible for learning a feature vector at the graph level, and (2) a weighted consensus clustering mechanism that operates on the learned graph embeddings using an ensemble of clustering predictors. The resulting clusters are generated via a weighted consensus method that takes into account the performance of every single predictor.

\textbf{Graph Embedding.} The goal is to learn a graph representation able to preserve the graph structure and code semantics. The representations learned by minimizing Equation~\ref{loss_f} are not suitable for this task since they only contain information that is helpful for classification and would filter out other information that is important for capturing the characteristics of the entire graph. 
Therefore, we consider solutions based on the graph autoencoder (GAE)~\cite{salha2019keep, kipf2016variational}, which consists of an encoder that learns a hidden representation and a decoder that can reconstruct the entire graph from this representation. Our implementation of this model involves utilizing a graph convolutional network (GCN) as the encoder and a simple inner product as the decoder.

In particular, we calculate graph embedding $Z$ and the reconstructed adjacency matrix $\hat{A}$ as follows:
\begin{align}
\hat{A} = \sigma (ZZ^{\top}),\ Z = GCN (X,A)
\end{align}
where $\sigma(\cdot)$ is the sigmoid function. We use binary cross entropy loss for reconstruction loss, which is used to train both the encoder and decoder. 
\begin{align}
\mathcal{L}_\mathit{recon}= \mathit{-  \sum_{i,j} \left[ A_{ij}log(\hat{A}_{ij}) +   (1-A_{ij})log (1-\hat{A}_{ij}) \right] }
\end{align}
where $\mathit{A_{ij}}$ is 1 if there is an edge between node $i$ and $j$, and 0 otherwise.  %After training the graph autoencoder using the reconstruction loss $\mathcal{L}_\mathit{recon}$ on graph $G$, denoted as $G = (X, A)$, we obtain the graph representation $Z$ from the encoder: $Z = \text{GCN}(X, A)$. Subsequently, we retrain the GAE from scratch for the next graph to obtain its corresponding graph representation.
After training the graph autoencoder with $\mathcal{L}_\mathit{recon}$ on graph $G = (X, A)$, we derive the graph representation $Z$ from the encoder: $Z = GCN(X, A)$, and then retrain the graph autoencoder from scratch for the next graph to get its graph representation.

\textbf{Weighted Consensus Clustering.} After obtaining an embedding for each graph, the next step is to assign a new label to each graph based on unsupervised clustering. In this paper, we introduce a novel weighted consensus clustering approach that leverages multiple clustering algorithms and assigns more weight to predictors that yield better clustering results.

First, we apply $P$  clustering algorithms to the graph embeddings, resulting in $P$ individual clustering solutions. Each solution assigns every graph to a specific cluster. In total, we have $P$ label solutions plus the original label if available. Next, we initialize a consensus matrix $\mathit{CM}$  with zeros. We iterate through all the solutions and update the consensus matrix on the fly. For every pair of graphs $i$ and $j$, if $i$ and $j$ belong to the same cluster, we update the matrix $\mathit{CM}$ using the formula:
\begin{align}
\mathit{CM[i][j] \mathrel{+}= 1 \times \frac{(s - (-1))}{2}}
\end{align}
Here, $s$ denotes the silhouette coefficient of the current clustering solution, which is the mean of the silhouette coefficient for all samples. %The silhouette coefficient for one sample is given as $\mathit{\frac{b-a}{max((a,b))}}$~\cite{rousseeuw1987silhouettes}, where $a$ is the mean distance between a sample and all the others in the same cluster and $b$ is the mean distance between a sample and all other samples in the nearest cluster. 
The averaged silhouette coefficient ranges from $-1$ for incorrect clustering to $+1$ for well-separated clustering. A superior clustering solution, characterized by a higher silhouette score, exerts a higher coefficient ($s$), hence a greater impact on the $\mathit{CM}$ matrix. Finally, we apply a clustering algorithm to the normalized matrix $\mathit{CM}$ to derive the final clusters, assigning each graph a new cluster label.

% \SetKwInput{KwInput}{Input}                % Set the Input
% \SetKwInput{KwOutput}{Output} 

% \begin{algorithm}[!ht]
% \DontPrintSemicolon
  
%   \KwInput{Your Input}
%   \KwOutput{Your output}
 
%   $\sum_{i=1}^{\infty} := 0$ \tcp*{this is a comment}
%   \tcc{Now this is an if...else conditional loop}
%   \If{Condition 1}
%     {
%         Do something    \tcp*{this is another comment}
%         \If{sub-Condition}
%         {Do a lot}
%     }
%     \ElseIf{Condition 2}
%     {
%     	Do Otherwise \;
%         \tcc{Now this is a for loop}
%         \For{sequence}    
%         { 
%         	loop instructions
%         }
%     }
%     \Else
%     {
%     	Do the rest
%     }
    
%     \tcc{Now this is a While loop}
%    \While{Condition}
%    {
%    		Do something\;
%    }

% \caption{Example code}
% \end{algorithm}
\section{Evaluation: Research Malware Dataset}

 In this section, we evaluate our approach using a popular Windows malware benchmark: the Microsoft malware classification challenge (Big-15).  We focus the evaluation on the detection of previously unseen malware families. Our assessment is conducted in two scenarios: one with the original labels and another using the labels generated by our graph cluster algorithm.  We evaluate different malware representations and current retraining approaches for each representation in Section~\ref{representation}.  We also test our approach on a real-world malware dataset collected from MalwareBazaar's malware database in 2024 (see Section~\ref{realworld}). Finally, we evaluate the family-level classification (see Section~\ref{exp_multi}). 
 
 %(2) How do different representations handle concept drift situation when employing non-adaptive techniques, other adaptation methods, and our proposed adaptation technique, when the number of training samples is limited? We introduce and justify these research questions to guide our evaluation of the approach. For each question, we outline one or more experiments and present the corresponding results. Before delving into them, we describe the experimental setup common to both questions. 

% \subsection{Experimental setup}\label{setup}
% We implemented our model using TensorFlow and Spektral, a library for GNN. The generator has 3 \verb"GIN" layers and ends with a global average pooling, and a dense layer with 256 neurons. The architecture of the classifier consists of 2 fully connected layers (\verb"FC_1",  \verb"FC_OUT"). The number of neurons in \verb"FC1" is 256. \verb"FC_OUT" is the output layers for label prediction. The discriminator has two layers with 256 hidden units and is followed by the softmax layer for domain prediction. Batch normalization is applied on every hidden layer. For training the model, we use the Adam optimizer with a learning rate of $\mathit{1e-3}$ for 60 epochs. The batch size is 32. The coefficient of the loss $\mathcal{L}_g$ is set to 0.1 , which allows the discriminator to be less sensitive to noisy signals during training. We set $\lambda = 0.1$ in $\mathcal{L}_c$  since we have much more labeled data from the source. %The code and processed graph data can be found here\footnote{}

\subsection{Evaluation based on original labels of Big-15}\label{exp_origin}

\subsubsection{Dataset}
The Microsoft Malware Classification Challenge (Big-15)~\cite{ronen2018microsoft} is one of the most used benchmarks for testing malware classification methods. In total, it has $21,741 $ malware samples where $10,868$ samples are labeled. Those labeled samples are from nine different malware families, namely Ramnit, Lollipop, Kelihos\_ver3, Vundo, Simda, Tracur, Kelihos\_ver1, Obfuscator.ACY, Gatak.  %Each sample consists of a .byte file, a binary representation of a hexadecimal number, and an .asm file, the disassembly outputs of IDA Pro. Microsoft removed the PE header to ensure file sterility.
In this experiment, we created CFGs from $10,868$ samples. %We iteratively select each of the malware family to be  the ``unseen family" in the target domain and the rest of families to be the ``exisiting families" in the source domain. We will discuss the experiment setup details in Section~\ref{st_setup_big_15} 

% Since Big-15 does not have benign samples,  we select two different benign binary datasets: BinaryCorp-3M~\cite{wang2022jtrans} from which will be used as the benign data for source domain and  Windows PE binaries which will be used as the benign data for target domain.  Note benign software can have concept drift as well so to have a realistic evaluation we use a benign dataset from a different operating system in the target. 

For the collection of Windows PE benign samples, as there is no dataset of benign binaries available due to copyright issues, we adopt the commonly used collection method~\cite{ling2024wolf}. We utilized virtual machines with clean installations of Windows 11, 10, and 8, along with common Windows applications. The resulting PE files were collected as benign samples, yielding a total of $16,000$ samples.

%Since Big-15 does not have benign samples,  we select two different benign binary datasets: BinaryCorp-3M~\cite{wang2022jtrans} and benign Windows PE files. BinaryCorp-3M consists of binaries from the official ArchLinux packages and Arch User Repository. The repository includes packages for editor, instant messenger, HTTP server, web browser, compiler, graphics library and cryptographic library. The dataset includes $8,357$ binaries compiled with gcc and g++ with different optimization levels. From the BinaryCorp-3M train, we obtained $6,392$ assembly files successfully using IDA Pro. % BinaryCorp offers a larger collection of 48,130 binaries. However, we opted not to use this larger version due to the size of Big-15, which contains 10,868 files. The reason behind this decision was to avoid any class bias that could arise from an imbalanced dataset. These 10,868 files are further divided into source and target malware data. In all our experiments, we maintain an approximate 1.4:1 ratio between malware and benign data in the source domain. 
%Benign Windows PE files are taken from installed folders of applications of legitimate software from different categories\footnote{They can be downloaded in https://download.cnet.com/windows/.}. In total, we successfully disassembled $1,000$ Windows PE files for our experiment. %The ``unseen” family from Big-15 will be combined with these files to form the target data, maintaining an approximate 1.5:1 malware to benign data ratio.

\subsubsection{Source and target datasets setup}\label{st_setup_big_15}

% The source and target domain samples need to be pre-annotated for training. This can be achieved with existing drift detection techniques -- not our focus. Our focus is to mitigate drifted samples and train a model that performs well on those samples with limited drifted labels. We thus designed our experiments using the "leave-one-out" method. We also tested on malwaredrift dataset, pre-divided into pre-drift and post-drift samples.  

We design our experiments using the ``leave-one-out" method to simulate an unseen malware family.  We pick one of the malware families as the previously unseen family (serving as the target malware data), whereas the remaining families serve as the source malware data.  We ensure the general applicability of our results by repeating this process with Ramnit, Lolipop, and Kelihos\_ver3, the dataset’s top three malware families.

Both the source and the target datasets contain benign samples.  We randomly split $8,000$ benign samples as the benign data for the source domain, while the target domain uses the rest of the $8,000$ benign samples. Ultimately, we successfully disassembled $6,510$ PE files from the source benign data, achieving an approximate malware-to-benign ratio of $1.2:1$ in the source dataset. The unseen family is combined with $5,768$ successfully disassembled benign Windows PE samples to form the target dataset, maintaining an approximate malware-to-benign data ratio of $0.35:1$.

% Both the source and the target datasets contains benign samples.   BinaryCorp-3M is used as the benign data for source domain, while the target domain uses benign Windows PE binaries.  We utilize distinct benign datasets in the source and target domains because benign software can have concept drift as well.  BinaryCorp provides a larger set of $48,130$ binaries. However, when we exclude one family, the source domain malware data will be around $8,000$. To prevent class bias from an extremely imbalanced dataset, we opt not to use the larger version of BinaryCorp. Instead, we use $6,392$ samples from BinaryCorp-3M as the benign data in the source dataset, achieving an approximate malware-to-benign ratio of $1.3:1$ in all experiments. The unseen family is combined with $1,000$ benign Windows applications to form the target dataset, maintaining an approximate malware-to-benign data ratio of $1.8:1$.

Since the Big-15 dataset does not provide sample timestamps, we split the source and target data conventionally.  The source and target datasets are randomly split into source training set ($75\%$), source testing set ($25\%$), target training set ($50\%$) and target testing set ($50\%$), respectively.  To prevent spatial bias~\cite{pendlebury2019tesseract}, the malware/benign class ratio is consistent across training and testing sets in all domains.  Our model is trained using the labeled source training set and a subset of the labeled target training set, then evaluated on the target testing set, which is unavailable during training. In continuous learning, a labeling budget limits the number of samples analysts can label.   We simulate this by randomly selecting $20, 50, 100, 200, 300, 500$ labels from the target training set and using them as the target training data.  In all experiments, the fraction of selected training samples is at a maximum of 50\% of the total target training set. We aim to detect malware in the target testing set with minimal labels, focusing on comparing model updating methods rather than improving performance through advanced sample selection. We chose random sampling, aware of its inefficiency, to demonstrate that if our method performs well with random sampling, it will excel with more sophisticated sampling techniques.

\subsubsection{Baselines}\label{big15-base}
We evaluate our scheme against existing graph-based malware classification methods and several enhanced baselines, which are adapted from previously published work with our improvements. This allows us to provide evidence that our method outperforms all these approaches.

%\ElisaText{Should we replace Current Methods with "Baseline Methods"?}
\textbf{Baseline Methods.} MAGIC~\cite{yan2019classifying}: Yan et al.~\cite{yan2019classifying} developed MAGIC, employing DGCNN, a type of GNN, to classify CFGs with basic blocks serving as nodes. MAGIC utilizes manually crafted token-level and block-level features to represent each basic block. Table~\ref{MAGIC} in the Appendix shows the features defined in MAGIC. 

MCBG~\cite{wu2021malware}: MCBG is another CFG-based malware classification model. It adopts a pre-trained BERT model to generate node embeddings instead of relying on handcrafted features. MCBG also adopts GIN as its classification model. 

MCBG and MAGIC are designed for supervised learning, assuming sufficient labeled data samples. Therefore, we employ cold-start learning (i.e., we train a classifier from scratch with target training labeled samples), which is consistent with past work.  We directly adopt their available public implementations with the default hyper-parameter. The details can be found in Appendix~\ref{app_baseline_big15}. 

%\ElisaText{If above we use "Baseline Methods", here we should use "Improved Baseline Methods".}
\textbf{Improved Baseline Methods.} Warm-start MAGIC and MCBG: We extend those approaches with warm-start learning. Rather than training a new model from scratch each time, we train the model with the source training data and continue training it with target training samples. The details of the implementation and retraining procedure are given in Appendix~\ref{app_baseline_big15}.%In addition, we found that freezing the weights of the initial layer of the old model performs better than retraining all the layers with the target samples on this dataset so we reported the higher results. The details of implementation can be found in Appendix~\ref{}.    
 
 DAN~\cite{long2015learning}: DAN was designed to generalize to test images different from those in the training set.  Specifically, it learns a domain-independent representation by reducing the discrepancy in domain distribution. This discrepancy is quantified using the Maximum Mean Discrepancy (MMD) loss. %, a kernel-based distance function that measures the difference between pairs of samples.
 %Our approach uses adversarial training to learn the domain invariant representation 
 We adapt DAN to support graph-based malware classification models. It learns a domain-independent graph representation via a shared feature extractor, which is based on GIN. Subsequently, we implement a classifier that uses this graph representation as input to determine if it is malware. The feature extractor is trained to minimize both the classification loss and the MMD loss, while the classifier is trained to minimize the classification loss. The MMD loss is computed using the hidden graph representations $h_G^s$ and $h_G^t$ derived from the feature extractor.  Following previous approaches, we opted for RBF as the kernel function. To ensure a fair comparison, the architecture of the feature extractor and classifier mirrors that of the generator and classifier components in our method.  See the Appendix for the details on the implementation of DAN~\ref{app_baseline_big15} and our approach~\ref{app_our}.  Both our approach and DAN aim to produce indistinguishable representations across domains. DAN uses MMD loss, a kernel-based distance function minimizing the disparity between the hidden representations of source and target samples, thereby achieving distribution matching. Our method is very different as it measures the disparity between distributions based on their separability by a neural network (discriminator). In image classification tasks, discriminator loss has shown superior performance compared to DAN~\cite{ganin2016domain}.

\begin{figure}[t]
  \centering
  \includegraphics[width=\linewidth]{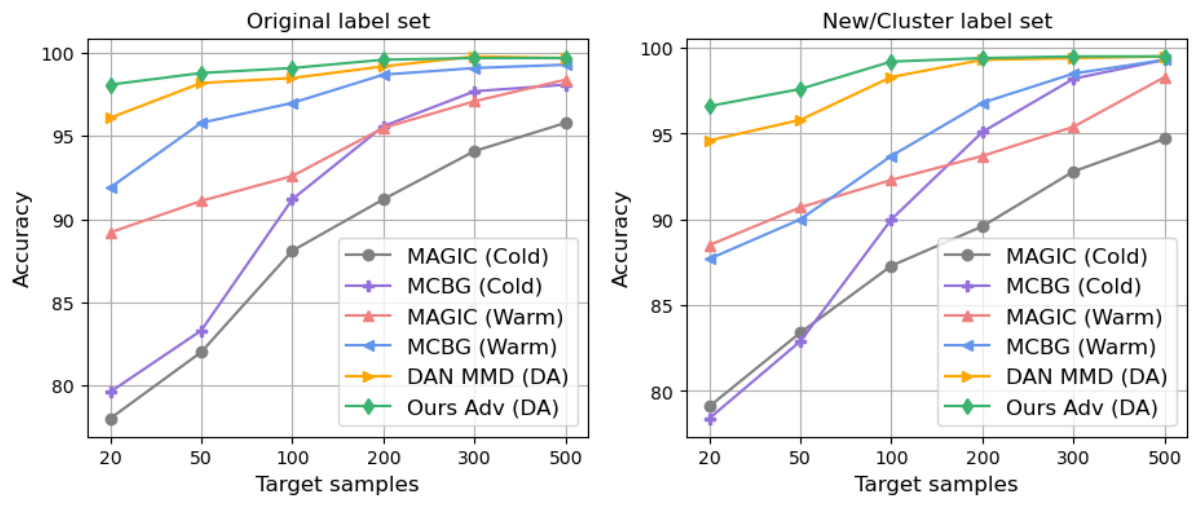}
  \caption{Given a set of fixed target training labels, we compute the accuracy of the target testing data for different baseline techniques and our method. The left diagram reports the averaged accuracy based on the original label set of Big-15, and the right one reports results based on the cluster label assignment.}%Visualization of the graph feature vector with their labels. The left one shows the data  with original labels from Big 15~\cite{ronen2018microsoft}, the right one shows the newly learned clusters. The legend represents the mapping between labels and colors.}
  \label{big15}
\end{figure}

\subsubsection{Results}
 %We evaluated how much our DA technique improves the performance of the classifier on the target testing data compared to all the baseline methods. We experiment with target training labels of 20, 50, 100, 200, 300, 500. 
 We run each experiment five times and compute the average results across all five iterations.  The accuracy of the three target families is shown in the top three graphs in Figure~\ref{app_big15} in the Appendix. The F1 score can be found in Table~\ref{app_big15_f1} in the Appendix. Here, we present the averaged results from the three target families in Figure~\ref{big15}. We observe that: 
 \begin{itemize}[leftmargin=*]
  \item  Our approach has the best accuracy in all experiments. It achieves over $98\%$ average accuracy when only trained with $20$ labeled samples from the target, demonstrating that our adaptation approach can effectively adapt to the new malware family with very few labeled samples.   
  \item DA methods (including DAN MMD and ours) perform better than the warm-start learning strategy, while the cold-start training yields the least effective results when trained with the same amount of target labels.  In the malware-detection task,  the discriminator loss demonstrates superior performance compared to DAN as well. 
  %\item  The warm-start learning strategy outperforms the cold-start one due to pre-training on a larger dataset with other malware families. 
  \item  MCBG (Warm) attains 91\%  accuracy with just $20$ samples. This is primarily due to the lack of clear separation among different family labels, which we further verify by the experiments presented in Section~\ref{exp_cluster}.
 \end{itemize}

In the next experiment, we initially demonstrate that the original labels of Big-15 are not well separated using common inter-label distance metrics, potentially resulting in an overestimation of the warm-start learning and DA methods. We create denser clusters for the same dataset using the graph-based clustering algorithm presented in Section~\ref{gca} and evaluate all the methods on this more challenging adaptation task.

\subsection{Evaluation based on cluster labels of Big-15}\label{exp_cluster}

\subsubsection{Inter-label distance}
We refer to three common metrics to measure the distance among labels: Silhouette Coefficient~\cite{rousseeuw1987silhouettes}, Calinski-Harabasz Index~\cite{calinski1974dendrite}, and Davies-Bouldin Index~\cite{davies1979cluster}. These metrics assess whether clusters are dense and well-separated. The Calinski-Harabasz Index scores higher for dense, well-separated clusters, whereas the Davies-Bouldin Index suggests better partitions when it approaches zero. These metrics are not directly applicable to graphs, so we use the method from Section~\ref{gca} to learn graph embeddings, compute metrics from transformed vectors and original labels, and present results in Table~\ref{metric}. All three metrics indicate poor separation of samples with different labels.  We also give a visual insight into the poor data partition based on the original label. The visualization is shown in Figure~\ref{tsne}.

% \usepackage{booktabs}
\begin{table}[t]
\caption{ Evaluation of the original labels and new clusters of Big-15 with distance-based metrics. All metrics confirm that our new clusters better separate data, with distant samples in different clusters and close samples in the same cluster.}
\label{metric}
\centering
\resizebox{0.8\linewidth}{!}{
\begin{tabular}{@{}lcc@{}}
\toprule
\textbf{Metric}                  & \multicolumn{1}{l}{\textbf{Original labels}} & \multicolumn{1}{l}{\textbf{New clusters}} \\ \midrule
Silhouette Coefficient  & -0.112                                       & -0.036                                    \\
Calinski-Harabasz Index & 0.540                                        & 61.150                                    \\
Davies-Bouldin Index    & 40.210                                       & 29.511                                    \\ \bottomrule
\end{tabular}
}
\end{table}

\subsubsection{Obtaining well-separated clusters}\label{obtain_cluster}
\begin{figure}[t]
  \centering
  \includegraphics[width=0.9\linewidth]{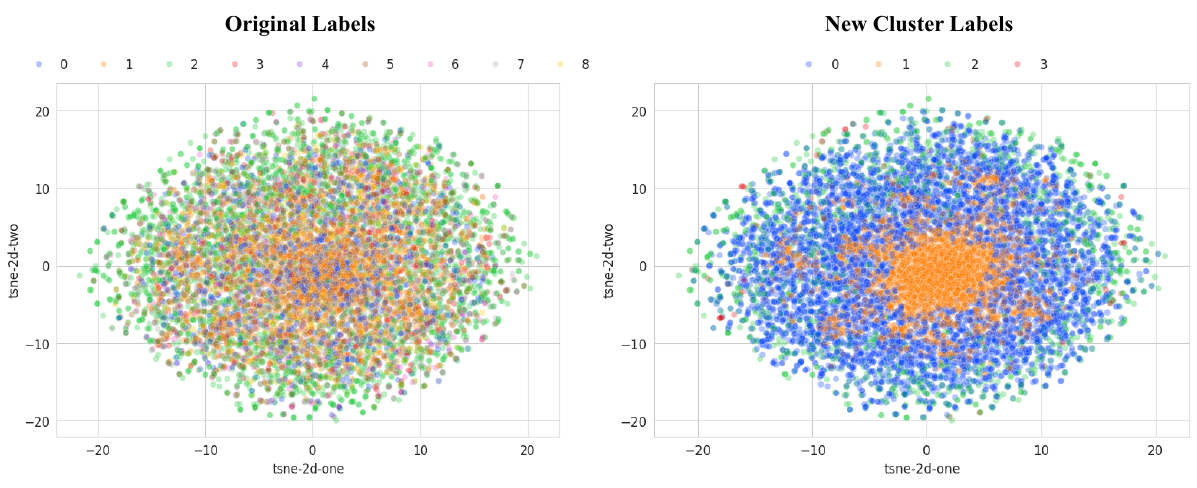}
  \caption{Visualization of the graph feature vector with their labels. The left part of the figure shows the data with the original labels from Big 15~\cite{ronen2018microsoft}, and the right one shows the newly learned clusters. The legend represents the mapping between labels and colors.}
  \label{tsne}
\end{figure}

We implemented the graph clustering algorithm in Section~\ref{gca} to obtain the new cluster labels for Big-15.  See Appendix~\ref{app_graph_clustering} for the implementation details. Figure~\ref{tsne} illustrates the new clusters, while Table~\ref{metric} presents the metric values for the new labels.  %All three metrics confirm that our new clusters better separate data. 

\subsubsection{Source and target datasets setup}\label{st_cluster_setup_big_15} We still employ the ``leave-one-out" strategy and pick one of the cluster labels as the ``unseen family" and the remaining clusters as the source malware data. We still use the same source benign data and target benign data. The train/test split ratio for each domain remains consistent with those detailed in Section~\ref{st_setup_big_15}.

\subsubsection{Results}

% %\ElisaText{The legend seems to be missing from the figure}
% \begin{figure}[t]
%   \centering
%   \includegraphics[width=\linewidth]{figures/ndss_mb.pdf}
%   \caption{Visualization of the graph feature vector with their labels. The left one shows the data  with original labels from Big-15~\cite{ronen2018microsoft}, the right one shows the newly learned clusters. The legend represents the mapping between labels and colors.}
%   \label{tsne}
% \end{figure}

We use the same baselines as in the previous experiments. %The malware data is associated with the new cluster labels.
Similarly, we conducted three adaptation tasks, each targeting a specific cluster (we have picked cluster 0, cluster 1 and cluster 2 as the target family).  We randomly sampled ${20, 50, 100, 200, 300, 500}$ labels from the target training set for each task. The full accuracy result and  F1 score are included in Figure~\ref{app_big15} and Table~\ref{app_big15_cluster_f1} in the Appendix.  We present
the averaged results from those three cases in Figure~\ref{big15}. To understand the impact of the more accurate labels on different training strategies,  we calculate the difference in the averaged accuracy of each approach based on the original label and cluster label and show the results in Table~\ref{avg_acc_diff} in the Appendix. The highlight results are:
\begin{itemize}[leftmargin=*]
    \item  Our approach maintains the highest accuracy across all experiments and demonstrates robust performance even in this more challenging adaptation task. It achieves an average accuracy of $96\%$ when trained with only $20$ labeled samples from the target domain.
    \item  In addition, the experiments show that the DA methods yield the highest performance, with warm-start learning trailing behind the DA methods and cold-start learning producing the lowest outcomes. %In a \ElisaText{what do you mean by "a repeated observation"? Perhaps you just say "In addition, the experiments show that the DA.."}repeated observation, the DA methods yield the highest performance, with warm-start learning trailing behind  \ElisaText{"trailing behind" what? Our approach?} and cold-start learning producing the lowest outcomes.
    \item MCBG (Warm) experienced the most substantial performance decline due to a more pronounced distribution shift between the source and target domains. When assessed on actual drifted labels, MCBG (Warm) saw an accuracy decrease of at most $5.8\%$. Both DAN and MAGIC (Warm) only experienced an average accuracy decrease of $0.8\%$ and $0.9\%$, respectively. Our method experienced the smallest drop of $0.5\%$ on average. %Cold-training performs similarly since it only sees the target data, but not the source data.
\end{itemize}

\section{Evaluation: Impact of Different Representations}\label{representation}
In this section, we conduct a comprehensive analysis of the effectiveness of various combinations of malware representation with diverse model update strategies. We consider the existing cold-start and warm-start learning approaches, as well as novel approaches grounded in DA. Our objective is to answer the following research questions:  

\textbf{Q1}: How effective is our adaptation technique when dealing with different malware representations, such as those that are content-based or image-based? 
%\ElisaText{I have slightly changed this sentence. Check that the new version is correct.}
Is our method the best for updating the model using these two representations?

\textbf{Q2}: Does our graph representation contribute to the overall model's performance? If so, to what extent?

\textbf{Q3}: In terms of cold-start and warm-start learning, which form of malware representation is more effective?

To answer those questions, we use the Big-15 dataset and the Windows benign dataset, obtain three representations (content-based, image-based, graph-based) for each malware sample, and test with various baselines for each representation.  For the malware data, we use the new cluster label as it poses a more challenging adaptation task.  The configuration of the source and target datasets adheres to the same process outlined in ~\ref{st_cluster_setup_big_15} and remains consistent across all representations. In what follows, we first describe the representation we consider. Then, we discuss the implemented baselines for each representation, followed by the results of our evaluation.

%in Section~\ref{results_features}. 

\subsubsection{Content-based features and baselines}

We select nine families of static analysis features that were curated in prior research works~\cite{aghakhani2020malware,ahmadi2016novel}. In total, we extracted a total of $965$ individual features from the assembly code sample in our datasets. A summary of the selected features is included in Appendix~\ref{details_content_features}.

%\ElisaText{Should we use Baseline here instead of "current"?}
\textbf{Baseline Methods.} SVM and MLP in both cold and warm settings: In continuous learning for malware detection, SVM and MLP are commonly used classifier models to learn from tabular features extracted from malware samples~\cite{yang2021cade, chen2023continuous}. Chen et al.~\cite{chen2023continuous} further enhanced this approach by incorporating warm-start training, which they found to be superior to cold-start for malware detection models.  The details of each classifier can be found in Appendix~\ref{implementation_content}.

\textbf{Improved Baseline Methods.} DAN: We modify DAN to accommodate MLP-based models for malware classification. Both the feature extractor and the classifier are implemented as multi-layered feedforward neural networks, and their combined structure mirrors the layer configuration of the baseline MLP. The feature extractor is trained to minimize both the classification and MMD losses, while the classifier’s training focuses on minimizing the classification loss. The computation of the MMD loss involves the hidden representations obtained from the feature extractor. Consistent with prior research, we choose RBF for kernel functions.
 
For our adaptation method, to maintain a fair comparison, we designed the architecture of the generator and classifier to match the topology of the feature extractor and classifier modules in DAN.  The key distinction is that our domain representation is acquired via adversarial learning loss in Equation~\ref{loss_f} and \ref{eq7}, and not through MMD loss. See Appendix~\ref{implementation_content} for details on the implementation of DAN and our approach.

\subsubsection{Image-based features and baselines} 
We adopt the same approach as in~\cite{vasan2020imcfn, ma2021comprehensive, bhardwaj2024overcoming} to transform a malware binary into an image.  This approach does not require any feature engineering and domain expert knowledge. It reads a given binary as a vector of 8-bit unsigned integers and then converts a vector into a 2D array. The height of the malware image is allowed to vary depending on the ﬁle size, and the
width of the malware image is ﬁxed. We apply this process to our datasets to obtain images of binaries.

\textbf{Baseline Methods.} Cold-start ResNet-50: We follow the previous works, which use several deep convolutional neural networks for malware prediction. We choose to implement the network components as ResNets-50 with short-cut connections since it shows the best prediction performance with images of malware~\cite{ma2021comprehensive}. In our first baseline, we train a ResNet-50 model from scratch using only target training samples, following approaches in~\cite{nataraj2011malware, singh2019malware} where no adaptation is involved. The implementation and training details can be found in Appendix~\ref{implementation_image}.  %The original ResNet-50's output layer contains 1000 neurons. As our task focuses on predicting whether a sample is malware or benign software, we modified the model by removing its last layer, adding a global average pooling layer, incorporating a fully connected layer with 256 neurons, and appending an output layer with 2 neurons.   

Warm-start ResNet-50:  This approach aligns with the predominant methodologies employed in classifying malware images with concept drift~\cite{ma2021comprehensive,vasan2020imcfn, bhodia2019transfer,kumar2021mcft}. Specifically, we utilize the same customized ResNet-50 architecture described earlier but initialize the model with weights from pre-training on ImageNet~\cite{deng2009imagenet}. Subsequently, we retrain the model using both source and target image data. This strategy not only leverages the network's knowledge of general images but also integrates insights learned from the source malware data.

\textbf{Improved Baseline Methods.} DAN: Our implementation of DAN and our methodology utilize a simple multi-layered CNN without using any pre-trained models. This demonstrates that the DA techniques surpass both the cold-training and warm-training approaches, even when using smaller models trained from scratch. For an in-depth explanation of the implementation of DAN and our method, please refer to Appendix~\ref{implementation_image}.

% To implement our adaptation approach, we opted to utilize the ResNet-50 with imagenet weight initialization as our generator, where the global pooling and two dense layers are moved to the downstreaming classifier. The discriminator consists of a global average pooling layer,  a dense layer with 1024 neurons and an output layer for domain prediction. The hyperparameters are the same with those outlined in Section \ref{setup}.            

\subsubsection{Results}\label{results_features}

\begin{figure}[t]
  \centering
 \includegraphics[width=0.9\linewidth]{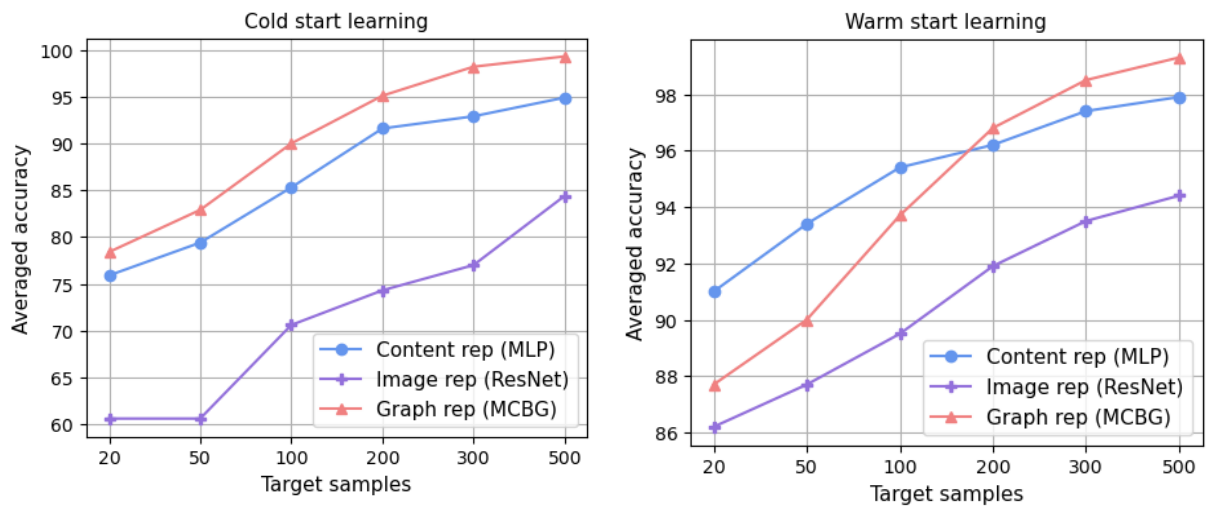}
  \caption{Comparison of different representations under cold-start learning (left) and warm-start learning (right).}
  \label{warmcold}
\end{figure}

%For each representation, the malware data is associated with the new cluster labels. 
Each baseline is assessed across three adaptation tasks, with each task leaving one cluster out as the target, and then we report the average from three tasks. The averaged accuracy of all feature representations alongside their respective baselines is reported in Table~\ref{Big-15-representation}. The F1 score can be found in Table~\ref{app-representation} in the Appendix. We have omitted the cold and warm strategies from the table for the graph representations as they are shown in Figure~\ref{big15}. Instead, we have selected the top-performing method from each representation under both cold-training and warm-training settings for a direct comparison in Figure~\ref{warmcold}. The following are our observations in response to the initial questions:

\begin{itemize}[leftmargin=*]
    \item Our shift adaptation component consistently has the highest accuracy across all feature representations, demonstrating the versatility of our adversarial DA technique with different malware representations (\textbf{Q1}).
    \item  All the key components in our pipeline contribute to the overall model's performance. Combining the adaptation approach with our graph representations achieves the best results. (\textbf{Q2}). 
    \item In cold-start training,  content-based and graph-based representations have similar performance, while the ResNet-50 image model using image features performs poorly, indicating the challenge posed by inadequate training data for large-scale neural network models. (\textbf{Q3}).
    \item In warm-start training, the image model shows the most significant improvement, and the graph representation surpasses the content-based features beyond $200 $ target samples (\textbf{Q3}).
\end{itemize}

% \usepackage{booktabs}
% \usepackage{multirow}
\begin{table}[t]
\caption{Averaged accuracy of baselines with various malware representations. For the content and image representation, we report the percentage change in accuracy from our adversarial (Adv) DA method to the peak performance among all baselines within the same representation. Ultimately, we demonstrate the improvement of our full pipeline (graph + Adv DA) compared to both content + Adv DA and image + Adv DA.}
\label{Big-15-representation}
\centering
\resizebox{0.95\linewidth}{!}{
\begin{tabular}{@{}cccrrrrrr@{}}
\toprule
\multicolumn{1}{l}{}                                                                                            &                                             &                                                         & \multicolumn{6}{c}{\textbf{Target samples}}                                                                                                                                                                                                                                                     \\ \cmidrule(l){4-9} 
\multicolumn{1}{l}{\multirow{-2}{*}{\textbf{\begin{tabular}[c]{@{}l@{}}Malware\\ Representation\end{tabular}}}} & \multirow{-2}{*}{\textbf{Strategy}}         & \multirow{-2}{*}{\textbf{Method}}                       & \multicolumn{1}{c}{20}                        & \multicolumn{1}{c}{50}                        & \multicolumn{1}{c}{100}                       & \multicolumn{1}{c}{200}                       & \multicolumn{1}{c}{300}                       & \multicolumn{1}{c}{500}                         \\ \midrule
\multicolumn{1}{c|}{}                                                                                           & \multicolumn{1}{c|}{}                       & \multicolumn{1}{c|}{SVM}                                & 67.2                                          & 71.4                                          & 75.4                                          & 79.3                                          & 83                                            & 85.6                                            \\ \cmidrule(lr){3-3}
\multicolumn{1}{c|}{}                                                                                           & \multicolumn{1}{c|}{\multirow{-2}{*}{Cold}} & \multicolumn{1}{c|}{MLP}                                & 75.9                                          & 79.4                                          & 85.3                                          & 91.6                                          & 92.9                                          & 94.9                                            \\ \cmidrule(lr){2-3}
\multicolumn{1}{c|}{}                                                                                           & \multicolumn{1}{c|}{}                       & \multicolumn{1}{c|}{SVM}                                & 70.2                                          & 74                                            & 79.1                                          & 83.2                                          & 86.7                                          & 90.5                                            \\ \cmidrule(lr){3-3}
\multicolumn{1}{c|}{}                                                                                           & \multicolumn{1}{c|}{\multirow{-2}{*}{Warm}} & \multicolumn{1}{c|}{MLP}                                & \textbf{91}                                   & 93.4                                          & 95.4                                          & 96.2                                          & \textbf{97.4}                                 & \textbf{97.9}                                   \\ \cmidrule(lr){2-3}
\multicolumn{1}{c|}{}                                                                                           & \multicolumn{1}{c|}{}                       & \multicolumn{1}{c|}{DAN (MMD) + MLP}                    & 90.8                                          & \textbf{93.9}                                 & \textbf{95.4}                                 & \textbf{96.9}                                 & 97.1                                          & 97.5                                            \\ \cmidrule(lr){3-3}
\multicolumn{1}{c|}{}                                                                                           & \multicolumn{1}{c|}{}                       & \multicolumn{1}{c|}{}                                   & \textbf{94.1}                                 & \textbf{95.2}                                 & \textbf{96.2}                                 & \textbf{97.1}                                 & \textbf{97.7}                                 & \textbf{97.8}                                   \\
\multicolumn{1}{c|}{\multirow{-7}{*}{\begin{tabular}[c]{@{}c@{}}Content-based\\ (CB)\end{tabular}}}             & \multicolumn{1}{c|}{\multirow{-3}{*}{DA}}   & \multicolumn{1}{c|}{\multirow{-2}{*}{Ours (Adv) + MLP}} & {\color[HTML]{32CB00} \textbf{$\uparrow$3.1}} & {\color[HTML]{32CB00} \textbf{$\uparrow$1.3}} & {\color[HTML]{32CB00} \textbf{$\uparrow$0.8}} & {\color[HTML]{32CB00} \textbf{$\uparrow$0.2}} & {\color[HTML]{32CB00} \textbf{$\uparrow$0.3}} & {\color[HTML]{FE0000} \textbf{$\downarrow$0.1}} \\ \midrule
\multicolumn{1}{c|}{}                                                                                           & \multicolumn{1}{c|}{Cold}                   & \multicolumn{1}{c|}{ResNet-50}                          & 60.6                                          & 60.6                                          & 70.6                                          & 74.3                                          & 77                                            & 84.4                                            \\ \cmidrule(lr){2-3}
\multicolumn{1}{c|}{}                                                                                           & \multicolumn{1}{c|}{Warm}                   & \multicolumn{1}{c|}{ResNet-50}                          & \textbf{86.2}                                 & \textbf{87.7}                                 & 89.5                                          & \textbf{91.9}                                 & \textbf{93.5}                                 & \textbf{94.4}                                   \\ \cmidrule(lr){2-3}
\multicolumn{1}{c|}{}                                                                                           & \multicolumn{1}{c|}{}                       & \multicolumn{1}{c|}{DAN (MMD) + CNN}                    & 85.4                                          & 87.5                                          & \textbf{89.9}                                 & 91.6                                          & 93.1                                          & 94                                              \\ \cmidrule(lr){3-3}
\multicolumn{1}{c|}{}                                                                                           & \multicolumn{1}{c|}{}                       & \multicolumn{1}{c|}{}                                   & \textbf{88.6}                                 & \textbf{90}                                   & \textbf{92.2}                                 & \textbf{93.1}                                 & \textbf{94}                                   & \textbf{94.6}                                   \\
\multicolumn{1}{c|}{\multirow{-5}{*}{\begin{tabular}[c]{@{}c@{}}Image-based\\ (IB)\end{tabular}}}               & \multicolumn{1}{c|}{\multirow{-3}{*}{DA}}   & \multicolumn{1}{c|}{\multirow{-2}{*}{Ours (Adv) + CNN}} & {\color[HTML]{32CB00} \textbf{$\uparrow$2.4}} & {\color[HTML]{32CB00} \textbf{$\uparrow$2.3}} & {\color[HTML]{32CB00} \textbf{$\uparrow$2.3}} & {\color[HTML]{32CB00} \textbf{$\uparrow$1.2}} & {\color[HTML]{32CB00} \textbf{$\uparrow$0.5}} & {\color[HTML]{32CB00} \textbf{$\uparrow$0.2}}   \\ \midrule
\multicolumn{1}{c|}{\begin{tabular}[c]{@{}c@{}}Graph-based\\ (GB)\end{tabular}}                                 & \multicolumn{1}{c|}{DA}                     & \multicolumn{1}{c|}{Ours (Adv) + GIN}                   & \textbf{96.6}                                 & \textbf{97.6}                                 & \textbf{99.2}                                 & \textbf{99.4}                                 & \textbf{99.5}                                 & \textbf{99.5}                                   \\ \midrule
\multicolumn{3}{c}{\textbf{Improvement over CB: Ours (Adv)}}                                                                                                                                                            & {\color[HTML]{32CB00} \textbf{$\uparrow$2.5}} & {\color[HTML]{32CB00} \textbf{$\uparrow$2.4}} & {\color[HTML]{32CB00} \textbf{$\uparrow$3}}   & {\color[HTML]{32CB00} \textbf{$\uparrow$2.3}} & {\color[HTML]{32CB00} \textbf{$\uparrow$1.8}} & {\color[HTML]{32CB00} \textbf{$\uparrow$1.7}}   \\
\multicolumn{3}{c}{\textbf{Improvement over IB: Ours (Adv)}}                                                                                                                                                            & {\color[HTML]{32CB00} \textbf{$\uparrow$8}}   & {\color[HTML]{32CB00} \textbf{$\uparrow$7.6}} & {\color[HTML]{32CB00} \textbf{$\uparrow$7}}   & {\color[HTML]{32CB00} \textbf{$\uparrow$6.3}} & {\color[HTML]{32CB00} \textbf{$\uparrow$5.5}} & {\color[HTML]{32CB00} \textbf{$\uparrow$4.9}}   \\ \bottomrule
\end{tabular}
}
\end{table}

\section{Evaluation: Real-world Malware Dataset}\label{realworld}

In this section, we first show the evaluation of our method on a more recent real-world malware dataset with diverse families using data split based on temporal information. Then, we show results on the effectiveness of our approach with obfuscated malware samples. 

\subsection{Detection performance on most recent malware samples}

\subsubsection{Dataset} The experiment aims to evaluate our method in a realistic setup using recent malware samples that reflect current trends. Each sample includes a timestamp to eliminate temporal bias~\cite{pendlebury2019tesseract}. Furthermore, rather than employing a ``leave-one-out'' approach, the drifting samples must be derived from diverse malware families.

Existing Windows malware datasets are outdated and do not meet our criteria. Unlike Android datasets like AndroidZoo~\cite{Allix:2016:ACM:2901739.2903508}, there is no  Windows malware dataset with original samples of both malware and goodware over the same period. The BODMAS dataset~\cite{bodmas} provides only malware binaries from five years ago, and the EMBER dataset~\cite{2018arXiv180404637A} only includes features from PE files, both malicious and benign, from seven years ago. EMBER’s original samples are only accessible via VirusTotal and are not publicly available for download.

Therefore, we collected a new malware dataset from March 2024 to September 2024 using the MalwareBazaar daily feed\footnote{https://bazaar.abuse.ch/api/}. After filtering non-PE samples and removing noisy samples with inconsistent labels, we created the MB-24 dataset. Detailed information, including the number of families and samples per month, is provided in Table~\ref{mb24_summary}. We use the collected $16,000$ benign Windows PE files as the benign dataset.

% \usepackage{booktabs}
\begin{table}[h]
\caption{Summary of the MB-24 dataset. }
\label{mb24_summary}
\centering
\resizebox{0.9\linewidth}{!}{
\begin{tabular}{@{}crrrrrr@{}}
\toprule
\textbf{Summary}    & \multicolumn{1}{l}{\textbf{Mar}} & \multicolumn{1}{c}{\textbf{April}} & \multicolumn{1}{c}{\textbf{May}} & \multicolumn{1}{c}{\textbf{July}} & \multicolumn{1}{c}{\textbf{Aug}} & \multicolumn{1}{c}{\textbf{Sep}} \\ \midrule
\# Samples          & 1505                             & 1080                               & 1496                             & 1618                              & 1613                             & 1337                             \\ \midrule
\# Malware Families & 104                              & 81                                 & 99                               & 126                               & 111                              & 92                               \\ \bottomrule
\end{tabular}
}
\end{table}

\subsubsection{Source and target datasets setup}

\begin{figure}[t]
  \centering
  \includegraphics[width=0.95\linewidth]{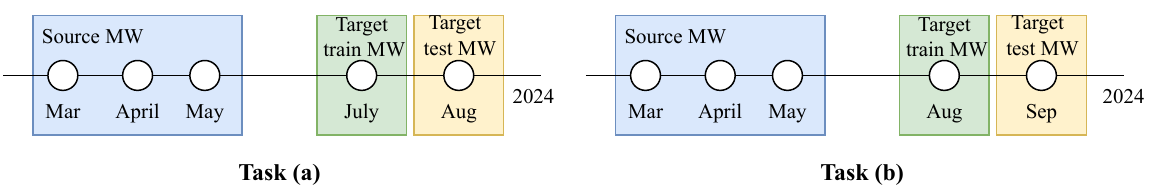}
  \caption{Source and target malware datasets setup for the monthly model update in July 2024 (Task (a)) and August 2024 (Task (b)).  }
  \label{mb24}
\end{figure}

We follow the time-consistent data split used in~\cite{chen2023continuous} to simulate the update of malware detection models in practice.  We use the data from March, April and May 2024
as the source malware set, the data in July 2024 as the target malware training set, and the data in August 2024 as the target testing set. In this way, we strictly follow the temporal constraint commonly used in the literature. We skipped June 2024 to ensure that malware distribution has drifted.  This is demonstrated in the left part of Figure~\ref{mb24}. We also tested another model update using the March-May 2024 data as the source malware set, the data in August 2024 as the target malware training set, and the data in September 2024 as the target testing set (right part of Figure~\ref{mb24}). 

 We randomly split $8000$ benign samples as the benign data for the source domain, while we evenly split the rest of $8000$ samples evenly to July, August and September\footnote{Since the benign PE files do not have timestamps, we randomly split the data conventionally.  A total of $6,510$ PE files were successfully disassembled from the source benign data. Additionally, about $1,922$ benign PE files were successfully disassembled for each of July, August, and September.}. In the end, the source dataset has a malware-to-benign ratio of $0.6:1$, and the malware-to-benign data ratio in the target training and testing sets (July, August and September) is 
 %\ElisaText{the wording "consistently approximately" does not sound good as using two adverbs one after the other does not sound good. Can we change the wording? For example can we say "is consistently about 0.8:1"?}
 consistently about $0.8:1$ to prevent spatial bias.

\begin{figure}[t]
  \centering
  \includegraphics[width=0.95\linewidth]{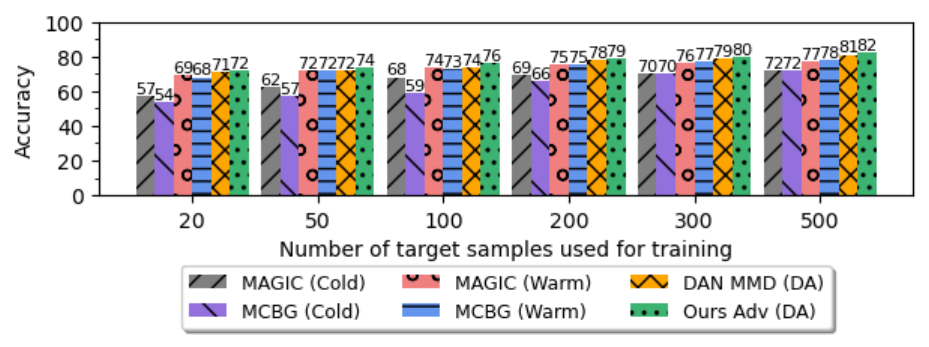}
  \caption{The averaged accuracy on the target testing data using the experimental setup described in Figure~\ref{mb24}. }
  \label{mb24_result}
\end{figure}

\subsubsection{Baselines} We use the same baselines of the previous experiments, with each baseline’s implementation detailed in Appendix~\ref{app_mb24_imple}. For each task, $20, 50, 100, 200, 300, 500$ labels are randomly sampled from the target training set to serve as the target training data. The random sampling is restricted to the target training set, with July data used for testing in August and August data used for testing in September. The primary focus of this experiment is to compare different model updating methods, hence the use of random sampling. Should the proposed method perform well with random sampling, its performance will further improve with more sophisticated sampling techniques such as uncertainty sampling~\cite{chen2023continuous} or the application of a rejection-threshold~\cite{barbero2022transcending, yang2021cade}. However, exploring these advanced techniques is beyond the scope of this work.

\subsubsection{Results}  We report the averaged accuracy of two adaptation tasks in Figure~\ref{mb24_result}. The F1 score can be found in Table~\ref{app-mb_24_result} in the Appendix. Our approach achieves the best results compared to all the baselines, followed by DAN and the warm-start training strategies.  All methods demonstrate a performance decline on the real-world dataset compared to the research dataset due to the complexity of the dataset. We conduct experiments to find the upper bound of prediction so we can reasonably assess the performance of our approach.  We train MCBG and MAGIC on August and September data separately, splitting each month’s data into a training set ($75\%$) and a testing set ($25\%$). The models are trained on the training set and evaluated on the test set of the same month. MCBG achieves an average accuracy of $77\%$ accuracy, while MAGIC achieves $79\%$. These results suggest that even if a human analyst labels $75\%$ of the samples in August ($1209/1613$) and September ($1002/1337$), the prediction accuracy would be the same as labeling only $200$ samples each month and using our approach. With the warm-start strategy, labeling $500$ samples achieves the same result.

 \subsection{Impact of obfuscation} To demonstrate the effectiveness of our approach against obfuscated malware, we conduct a two-step experiment.

First step:  We design the experiment using 
%\ElisaText{"The first experiment is tested" does not sound correct. Also I could not find task (b) in Figure 8.}
Task (b) in Figure~\ref{mb24} where the source malware data is from March to May, the target training malware is from August, and the target testing malware is from September.   We obfuscate all target testing malware samples with Hyperion\footnote{https://nullsecurity.net/tools/binary.html}, a runtime PE-Crypter.  The source malware and target training malware samples remain unobfuscated. Neither the baselines nor our model are exposed to obfuscated samples during training.  The goal is to compare the performance of each method trained with unobfuscated malware when testing on obfuscated malware. 

Second step: We show that accuracy improves by adding a few obfuscated samples to the target training set. Specifically, the model is trained using the source data, $100$ unobfuscated August samples, and either $10$ or $20$ obfuscated August samples, and then tested on the obfuscated September data.

\begin{figure}[h]
  \centering
  \includegraphics[width=0.95\linewidth]{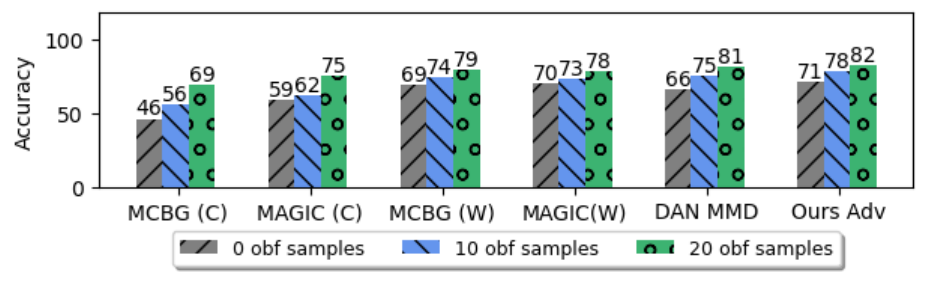}
  \caption{Accuracy improvement of different methods tested on September data by adding $10$ and $20$ obfuscated samples from August to the $100$ unobfuscated August target training data. }
  \label{mb24_obf}
\end{figure}

The results of the first experiment are reported in Figure~\ref{app_mb24_obf_ma} in the Appendix, while the results of the second experiment are presented in Figure~\ref{mb24_obf}. The first experiment shows that our approach exhibits the lowest accuracy degradation when tested on obfuscated malware samples not encountered during training. In the second experiment, incorporating $20$ obfuscated samples from August into the training set improves the accuracy of our approach by $11\%$ in predicting the obfuscated samples in September. Although this experiment represents an initial exploration of testing the approach with obfuscated malware, it provides insights for real-world deployment. In practice, the robustness of the model can be enhanced by obfuscating existing samples using various obfuscation tools and incorporating them into the training process. The experiment indicates that this strategy can significantly enhance protection against obfuscated samples.

\section{Evaluation: multi-family classification}\label{exp_multi}

In this section, we address malware classification at the family level. The classification process can be 
approached as a closed-set or an open-set DA problem. In the closed-set DA scenario, both the source and target domains contain identical malware families. Conversely, the open-set DA scenario allows for new malware families in the target domain that are not present in the source domain. We have conducted experiments under both closed and open-set conditions. Our findings demonstrate that our approach surpasses the current state-of-the-art methods in closed-set family-level classification and can be seamlessly adapted for open-set classification. Furthermore, we provide evidence supporting the superior performance of our approach, which effectively aligns samples from the same malware family across pre-drift and post-drift domains within the latent space. Additional details are provided in Section~\ref{evidence_drift}.

\subsection{Closed-set DA}

\subsubsection{Dataset}

In this experiment, we use the MalwareDrift dataset~\cite{ma2021comprehensive}\cite{wadkar2020detecting}. 
%, for evaluation purposes. 
The dataset encompasses samples from 7 distinct malware families, namely Bifrose, Creeinject, Obfuscator, Vbinject, Vobfus, Winwebsec, and Zegost, each representing various types of malware such as Trojans, worms, adware, and backdoors. Samples within each family are categorized into pre-drift and post-drift segments, with the partition determined at points of significant evolution. The evolution was determined using an ML-based tracking method that trains an SVM on malware families over time windows.  The $\chi^2$ statistic is employed to quantify variations in SVM weights over these sliding time windows. Significant spikes in the $\chi^2$ timeline indicate alterations in the malware's characteristics. Previous work by Ma et al.~\cite{ma2021comprehensive} has shown that models trained on the pre-drift data perform poorly on the post-drift data.  Figure~\ref{app_mb24_obf_ma} in the Appendix shows performance data from~\cite{ma2021comprehensive} on pre-drift and post-drift testing data for five models trained solely on pre-drift training data. The reduction in accuracy when tested on the post-drift data ranges from $23.0\%$ to $38.7\%$.  %It is thus clear that all considered models experience a substantial performance decrease when confronted with code evolution within the same malware families.
%The results are sourced from ~\cite{ma2021comprehensive}. 
%As we can see from the figure, the reduction in accuracy ranges from 23.0\% to 38.7\%. It is thus clear that all considered models experience a substantial performance decrease when confronted with code evolution within the same malware families.

%To show the performance decrease when the model predicts significantly changed malware from the time of training, we results on both pre-drift and post-drift data using five approaches trained solely on pre-drift data in Figure~\ref{source}. These results are sourced from the study by Ma et al.~\cite{ma2021comprehensive}. It is evident that all existing methods experience a substantial performance decrease when confronted with code evolution within the same malware families. The reduction in F1-score ranges from 23.0\% to 38.7\%.

% \begin{figure}[h]
%   \centering
%   \includesvg[width=0.8\linewidth]{../figures/malwaredrift.svg}
%   \caption{Accuracy of source only approach (a); Accuracy on post-drift data using 10-45 labeled data from each post-drift family (b).}
%   \label{source}
%   \vspace{-3mm}
% \end{figure}

\subsubsection{Source and target datasets setup} 

%Since the post-drift samples exhibit significant divergence compared to the pre-drift domain, our evaluation utilizes the original labels without the need to run our clustering algorithm for reassigning cluster labels. 

The source domain includes pre-drift data from seven malware families and benign Windows samples, while the target domain comprises post-drift data from the same seven families and different benign Windows samples. Our objective is to train a classifier to identify whether a sample from the target domain belongs to one of the eight classes (benign or one of the seven malware families) using a limited number of target samples. The post-drift data is ordered based on time and grouped by family. For the target training set, the first $10, 20, 30, 40, 45$ labeled samples from each class in the post-drift dataset are selected. The remaining samples form the target testing set.  This ensures that the target testing set maintains the same class ratio as the original post-drift dataset, preventing spatial bias~\cite{pendlebury2019tesseract}.   In all experiments, the fraction of selected target training samples is at a maximum of $19\%$ of the total post-drift dataset. All pre-drift data is used as the source training set. The source and target datasets are selected without temporal bias~\cite{pendlebury2019tesseract}, as all data in the training set precedes the testing data temporally.

\subsubsection{Baselines}

We use the same baselines approach listed in Section~\ref{big15-base}. We include the implementation details of each baseline in Appendix~\ref{implementation_drift}.%The architecture of the models and hyperparameter settings remain consistent with those outlined in Sections~\ref{big15-base} and in Appendix~\ref{}, except for the output layer of the classifier, which is modified to accommodate 7 classes.
\begin{figure}[t]
  \centering
  \includegraphics[width=0.95\linewidth]{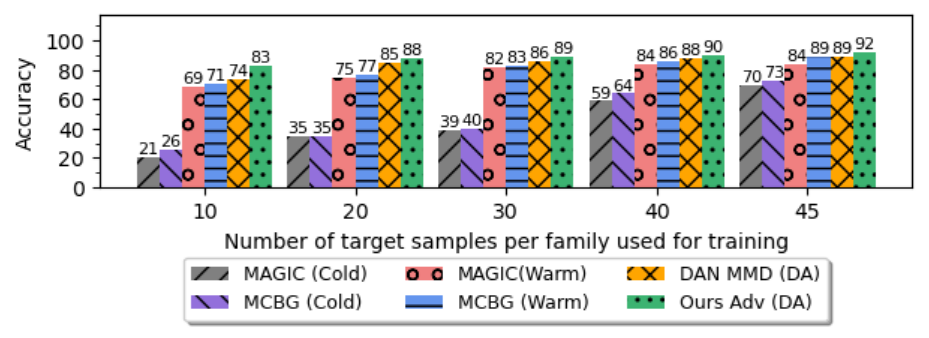}
  \caption{Accuracy on post-drift data using $10-45$ labeled data from each post-drift family. Table~\ref{app_multi} in the Appendix shows the F1-score.}
  \label{malwaredrift}
\end{figure}

\subsubsection{Results}

The results, presented in Figure~\ref{malwaredrift},  show that:
\begin{itemize}[leftmargin=*]
    \item Our method significantly outperforms other baseline methods in this closed-set family family classification task. It achieves an $83\%$ accuracy in classifying family labels for the post-drift set with just $10$ labeled samples per class. 
    \item In comparison to other baselines, DAN exhibits superior performance.  MAGIC (Cold) and MCBG (Cold) fail to converge due to insufficient data, especially for complex tasks like multi-class classification.  MAGIC (Warm) and MCBG (Warm), while not as effective as ours, have a considerable performance boost compared to when the models are trained in a cold-start setting.  
\end{itemize}

\subsection{Open-set DA}\label{openset}

In open-set classification, the target dataset includes unknown malware families that are not present in the source. There are state-of-the-art approaches for this open-set DA problem in the vision domain~\cite{panareda2017open, saito2018open}. Such approaches are able to (1) accurately classify target samples belonging to known classes and (2) detect and label unknown classes as ``unknown''. We have demonstrated the first capability with our closed-set experiment and now extend our approach to achieve the second capability in open-set malware family classification.

To handle unknown classes, we first ensure that these samples are not misclassified as benign, which is essential to prevent model evasion attacks. We then determine if the test sample is an outlier relative to the existing classes in the target training data, as outliers may signify new malware families. These outlier samples are set aside, and once a sufficient number is collected, they can be clustered using our graph-based approach. Human analysts would then typically review and assign new family labels to samples from each cluster, as commonly done in various existing approaches~\cite{liang2021fare}. These new classes can be used to update our model or to train additional classifiers as needed.

To detect new families as outliers or unknowns, we introduce an outlier detection module based on one-class SVM that seamlessly integrates with our trained model.  We train a one-class SVM on the learned latent graph representations of the post-drift training data and use the trained SVM to classify the latent representations from the test samples (we pass them to the generator to obtain such representations).
% \ElisaText{The two sentences below are a bit confusing. The SVM assigns a label of -1 to the samples that are outliers, and then we assign them an "unknown" label. Does it mean that "unknown" is another name for -1?}
% The SVM associates a label of (-1) to the sample if the test sample is an outlier. To the outlier samples, we assign an ``unknown" label to them.  Does this help clarify the confusion? 
The SVM assigns a label of ($-1$) to test samples identified as outliers. We then reassign an ``unknown'' label to these outlier samples.

We utilize the same source and target datasets setup as in the closed-set domain DA and train our model using the source training data along with 45 labeled samples from each class in the target training set. During testing, we provide three datasets, each containing $1$, $5$, and $9$ new families from the Big-15 dataset, respectively. Our approach is evaluated on these testing sets using the following metrics: \textbf{Evasion success rate:} The percentage of samples from unseen malware families misclassified as benign. \textbf{Outlier detection rate:} The percentage of samples from unseen malware families correctly identified as the ``unknown'' class.

\begin{figure}[h]
  \centering
  \includegraphics[width=0.6\linewidth]{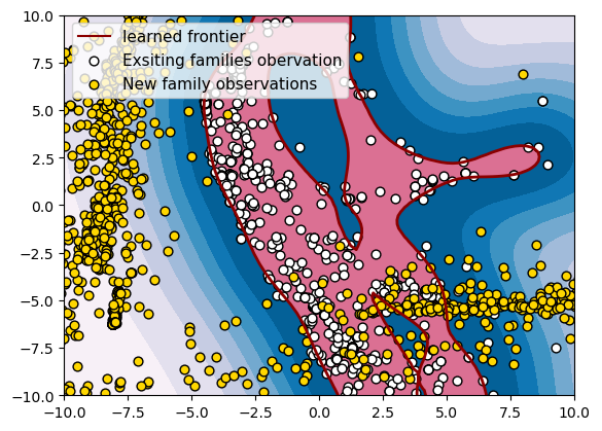}
  \caption{Visualization of the decision function learned by the outlier detection module: the new family observations are outside the learned frontier. }
  \label{frontier}
\end{figure}

The results of both metrics are presented in Table~\ref{outlier}. When our model is tested on new malware families, the evasion rate remains consistently low (below $5\%$), demonstrating robustness against evasion attempts. Additionally, the outlier detection module, trained on post-drift data features extracted by the generator, effectively identifies unknown samples. Figure~\ref{frontier} visualizes the decision boundary of the outlier detection module, with white dots representing latent representation from existing family observations and yellow dots representing latent representation from new family observations.

\begin{table}[h]
\caption{Evasion rate and detected unknown samples ratio on three testing datasets with 1, 5 and 9 new families. }
\label{outlier}
\centering
\resizebox{0.95\linewidth}{!}{
\begin{tabular}{@{}crrrrrr@{}}
\toprule
\textbf{Metrics}               & \textbf{1 new family} & \textbf{5 new families} & \textbf{9 new families} \\ \midrule
Evasion success rate (\%)              & 1.05                               & 4.04                                 & 4.46                             \\ \midrule
Detected/Total unknown samples & 2396/2476                          & 7662/9122                            & 8921/10711                       \\ \bottomrule
\end{tabular}
}
\end{table}

\subsection{Visualization of extracted features}\label{evidence_drift}
\begin{figure}[h]
  \centering
  \includegraphics[width=\linewidth]{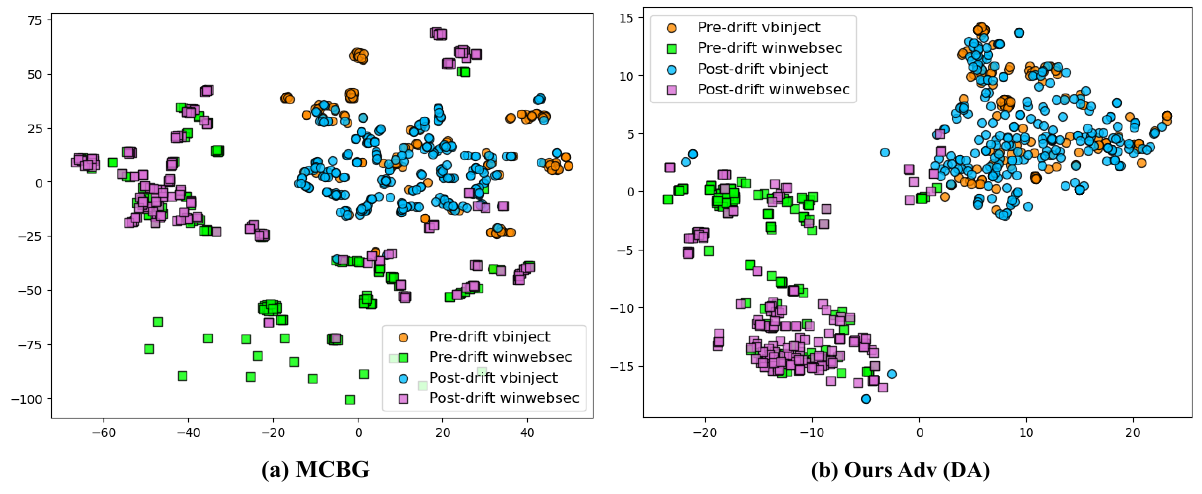}
  \caption{The effect of adaptation on the distribution of the extracted features from pre-drift and post-drift data (best viewed
in color). Each point represents a sample. Yellow and blue dots: pre-drift and post-drift ``vbinject". Purple and green boxes: pre-drift and post-drift ``winwebsec".   }
  \label{invariant}
\end{figure}

We show evidence that the features extracted by our generator are drift-invariant. Figure~\ref{invariant} shows the effect of DA on the distribution of latent-space features from two malware families in the MalwareDrift dataset. The t-SNE visualizations show the latent graph representation (learned features) from our adaptation model (right) and the features learned by the MCBG model trained only on pre-drift data (left). DA effectively reduces distribution divergence between pre-drift and post-drift data, as samples from the same malware family in both domains are well clustered together. The clear boundary between these clusters indicates that the learned representations are class-discriminative. Conversely, the MCBG model suffers from a distribution shift, with post-drift data points diverging from pre-drift data points within the same family. 

\section{Discussion}

\textbf{Computational Runtime.}  
%Empirically, 
We have recorded the average runtime for each step in our pipeline for the experiments on MB-24 (Section~\ref{realworld}), and the results are in Appendix~\ref{app_runtime}. Our training overhead is acceptable compared to the warm-start and cold-start training strategies. The majority of the pipeline time is dedicated to data extraction and transformation, which is typical for most ML-based methods in novel applications. Extracting content-based features is the most time-consuming because it involves processing assembly files, which can be several gigabytes in size, depending on the binary. Converting the binary to an image representation is the fastest among all the representation processes, but as we have shown in the experiments, the image representation performs the worst when the training data size is small. Our graph representation processing time is between the two, achieving the best or comparable results on the datasets we have used in our evaluation.  

\textbf{System Architecture.} In this paper, we focus on adapting malware for one system architecture or one instruction set. We carried out our evaluations on x86, given the notable absence of work on Windows malware.  However, our approaches also apply to other systems, such as Android malware. 

\textbf{Robustness of Graph Representation.} 
Ling et al.~\cite{ling2024wolf} demonstrated that most model evasion attacks on malware CFGs are ineffective. Effective attacks require complex transformations, such as call-based redividing transformation~\cite{ling2024wolf}. Adversarial training is the most effective defense against these attacks. To implement adversarial training, one can generate adversarial samples from various attacks and incorporate them into both domains. The training process and model remain unchanged, with the addition of new adversarial samples for each domain. Future work will further explore the integration of this approach with our DA method.

\section{Related work}\label{related_work}
%\ElisaText{It seems to me that this section just discusses related work. It does nor provide any background. So I would remove "Background" from the title and I would move the related work before the conclusions. Most papers do the same.}
\textbf{Supervised Malware Classification.}
The learning-based approaches for analyzing executable files fall into three main categories, namely learning from static features~\cite{ahmadi2016novel, aghakhani2020malware, nataraj2011malware, vasan2020imcfn, singh2019malware, bhodia2019transfer,yan2019classifying, someya2023fcgat},  dynamic behaviors~\cite{yin2023method, hou2023proteus}, or a combination of both~\cite{dambra2023decoding}.
Control Flow Graph (CFG), derived from assembly code, has shown exceptional efficacy in malware classification~\cite{yan2019classifying,wu2021malware}. %Yan et al.~\cite{yan2019classifying} developed MAGIC, employing DGCNN, a type of Graph Neural Network (GNN), to classify CFG with basic blocks serving as nodes. MAGIC utilizes manually crafted features to generate node embeddings for each basic block. Wu et al.~\cite{wu2021malware} proposed MCBG that trains a Graph Isomorphism Network (GIN) on CFGs. %MAGIC, MCBG, and FCGAT have not tackled concept drift issues in malware detection. 
Among them, MCBG is the closest to our approach in that it uses CFGs and GIN. However, MCBG is designed for supervised learning with the assumption of sufficient labeled samples. It lacks the capability for DA and has not been tested on drifted malware. In contrast, we address concept drift in malware classification, where drifted samples are typically scarce, by
%. We tackle this issue 
using adversarial DA on CFGs.  % Ma et al.~\cite{ma2021comprehensive} found no significant superiority among malware representations in a supervised setting. In this work, we focus on evaluating the performance of different representations with scarce new labels.  

\textbf{Malware Classification with Concept Drift.} 
The state-of-the-art solutions for mitigating concept drift employ a similar two-step approach: (1) Sample Selection: Identifying and manually labeling high-impact labels that are critical for learning the distribution of new malware. (2) Model Retraining: Updating the models with these newly labeled data. This procedure is repeated to ensure that the classifier keeps up to date with the latest malware. The primary difference between these solutions is the sample selection technique used. Jordaney et al.~\cite{jordaney2017transcend} and Barbero et al.~\cite{barbero2022transcending} proposed conformal evaluators that identify and reject examples that differ from the training distribution. %OWAD~\cite{han2023anomaly}  focuses on the shift of normal data as the models are trained with purely normal data and without any knowledge of the anomaly. 
Yan et al.~\cite{yan2019classifying} introduced a method called CADE to detect drifting samples based on contrastive representation learning. Chen et al.~\cite {chen2023continuous} proposed a novel uncertainty score and a pseudo loss for sample selection. For the second step, the models are updated using either cold-start or warm-start learning~\cite{chen2023continuous}. We focus on the second step, introducing a novel approach based on DA. Our experiments demonstrate that this approach outperforms existing retraining methods. We believe our work is orthogonal to the sample selection techniques and they could be incorporated into ours. Drift detection techniques could be used to select a minimal subset of data that diverges from the previous samples. Subsequently, these identified samples could serve as the post-drift data for training our model.

\textbf{Domain Adaptation.}  
%DA is a learning strategy developed to tackle the scarcity of labeled data. 
The principle behind DA is to leverage a labeled dataset from a source domain to assist in classifying data from a related yet distinct target domain that lacks labeled samples. %Transferring the classification capability from one domain to another is called Domain Adaptation (DA). 
Here, we briefly discuss two previous DA approaches closely related to ours.  The first approach focuses on aligning the statistical distribution shift between the source and target domains through a 
%certain mechanisms. The 
Domain Adversarial Network (DAN)~\cite{long2015learning}. A DAN utilizes the MMD loss to compare and reduce distribution shifts. The second approach focuses on learning a hidden representation using two rival networks: a generator and a discriminator. The Domain Adversarial Neural Network (DANN)~\cite{ganin2016domain} is one of the prominent methods of this kind. % employs three network components: a feature extractor, a label predictor, and a domain classifier. The generator is adversarially trained to increase the domain classifier’s loss by reversing its gradients. Simultaneously, the generator and label predictor are trained to create a representation that includes domain-invariant features for classification. %The Adversarial Discriminative Domain Adaptation (ADDA)~\cite{tzeng2017adversarial} method uses similar network components but involves multiple stages in training the model’s three components.
However, DAN and DANN have only been used and evaluated for image classification tasks. 
%on source and target domains of images. 

Some approaches were proposed for DA on graph-structured data. %CDNE~\cite{shen2020network} learns transferable node embeddings by minimizing the maximum mean discrepancy (MMD) loss. However, it cannot jointly model network structures and node attributes. 
AdaGCN~\cite{dai2022graph} is designed for node-level classification tasks and was evaluated on predicting paper topics in citation networks.  In our work, we address a graph-level classification problem for classifying malware with concept drift. This requires redesigning the losses and the model, leading to significant differences from prior works. Also, our solution includes a carefully designed pipeline with two other components to produce CFGs with high-quality embeddings. As shown in Section~\ref{representation}, the absence of integration of those CFGs results in worse performance with the same DA technique. %Our solution requires a carefully designed pipeline with two other components, producing high-quality embeddings for CFGs. %Our approach uses domain adaptation on control flow graphs of malware, which requires a novel pipeline with carefully design components for learning from complicated  code semantics and execution flow. 
Recently, DA has been used to address data scarcity in security functions like network intrusion~\cite{singla2020preparing} and vulnerability detection~\cite{li2023cross,zhang2023cpvd}. VulGDA~\cite{li2023cross} and CPVD~\cite{zhang2023cpvd} use DA for cross-project vulnerability detection at the source code level, transforming Code Property Graphs into vector features and learning transferable features with MMD loss and adversarial training, respectively. Our approach differs from VulGDA and CPVD in terms of purpose, operational level, and model input. VulGDA and CPVD work at the source code level for a different security function and cannot be adapted for malware detection, as most malware is in binary form. Also, both approaches utilize intermediate vector representations as the input, derived from training a GNN with source labels, which may introduce a source-only bias in the target vectors. In contrast, our approach employs graphs directly as inputs to the generator, thereby mitigating potential biases originating from the source domain.
%\ElisaText{Can we mention more differences compared with VulGDA and CVPD?}

% Recently, DA has been used to address the data scarcity issue in other security functions, such as network intrusion detection~\cite{singla2020preparing, li2019dart} and vulnerability detection~\cite{li2023cross,zhang2023cpvd}. VulGDA~\cite{li2023cross} and CPVD~\cite{zhang2023cpvd} employ DA to extract features characterizing vulnerabilities from graph representations of source code, enabling cross-project learning. They transform Code Property Graphs into vector features and then learn transferable features using 
% %\ElisaText{If the acronym MMD has not been introduced, please introduce it.}
% MMD loss and adversarial training, respectively. Both methods train a GNN with source labels to generate vector representations for both source and target domains. Our approach differs in that we focus on malware binaries. We use graphs directly as inputs to the generator instead of intermediate vector representations, avoiding potential biases from the source domain. 

\section{Conclusion}

This paper addresses the classification of drifted malware and the challenge of adapting models with limited labels. Our approach is based on adversarial domain adaptation. It operates on CFGs, exploiting the consistent characteristics of malware in terms of assembly code semantics and control flow execution.  Our process comprises three main elements: constructing CFGs, extracting vertex features, and adapting to drifted samples. %To retain the semantics of the code from the instructions, we utilize an assembly language model to generate high-quality feature embeddings for each instruction within every node.
We have extensively compared our training approach with approaches from previously published work or adapted to support malware classification.  
The experimental results show that our approach outperforms others in three distinct adaptation tasks with increasing adaptation complexity: evaluation on research datasets,  evaluation on the latest real-world malware samples, and classification of multiple malware families that have evolved. We also demonstrate that our adaptation component can be used with other malware representations to improve performance while our graph representation achieves the best results. We conclude that our approach can effectively improve the model performance when trained with scarce new labels.

\section*{Acknowledgment}
We thank Md Ajwad Akil from cyber2slab of Purdue
University for their valuable time and effort in collecting the
MB-24 dataset.  The work reported in this paper has been supported by the National Science Foundation (NSF) under Grants 2229876 and 2112471.



%\bibliographystyle{IEEEtranS}
\bibliographystyle{IEEEtranS}
\bibliography{main}

\appendices

\section{}
\subsection{Implementation details and additional results on Big-15}

\subsubsection{Our model implementation details}\label{app_our} 
We implemented our model using TensorFlow and Spektral, a library for GNN. The generator has $3$ \verb"GIN" layers and ends with a global average pooling, and a dense layer with $256$ neurons. The architecture of the classifier consists of $2$ fully connected layers (\verb"FC_1",  \verb"FC_OUT"). The number of neurons in \verb"FC1" is $256$. \verb"FC_OUT" is the output layer for label prediction. The discriminator has two layers with $256$ hidden units and is followed by the softmax layer for domain prediction. Batch normalization is applied in each hidden layer. For training the model, we use the Adam optimizer with a learning rate of $\mathit{1e-3}$ for $60$ epochs. The batch size is $16$. The coefficient of the loss $\mathcal{L}_g$ is set to $0.1$, which allows the discriminator to be less sensitive to noisy signals during training. We set $\lambda = 0.1$ in $\mathcal{L}_c$  since we have more labeled data from the source. %The code and processed graph data can be found here\footnote{}

\subsubsection{Baseline implementation details}\label{app_baseline_big15}

MAGIC (Cold): MAGIC utilizes handcrafted features to represent each basic block, which are listed in Table~\ref{MAGIC}. The classification model is DGCNN, which uses a SortPooling layer. Our DGCNN topology is based on the ones used in~\cite{yan2019classifying} to be comparable to previous work. We train the model using the Adam optimizer with a learning rate of $\mathit{1e-3}$ for $60$ epochs.

\begin{table}[h]
\caption{Basic block attributes defined in MAGIC}
\label{MAGIC}
\centering
\resizebox{0.5\linewidth}{!}{
\begin{tabular}{@{}cl@{}}
\toprule
\multicolumn{1}{l}{Attribute Type} & Attribute Description            \\ \midrule
\multirow{9}{*}{token-level}       & \# Numeric Constants             \\
                                   & \# Transfer Instructions         \\
                                   & \# Call Instructions             \\
                                   & \# Arithmetic Instruction        \\
                                   & \# Compare Instructions          \\
                                   & \# Mov Instructions              \\
                                   & \# Termination Instructions      \\
                                   & \# Data Declaration Instructions \\
                                   & \# Total Instructions            \\ \midrule
\multirow{2}{*}{block-level}       & \# Offspring, i.e., Degree       \\
                                   & \# Instructions in the Block     \\ \bottomrule
\end{tabular}
}
\end{table}

MCBG (Cold):   MCBG uses the GIN-JK model as the classifier. We use their default GIN-JK model with the same hyperparameters used in~\cite{wu2021malware}. MAGIC is trained using the Adam optimizer with a learning rate of $\mathit{1e-3}$ for $60$ epochs.

MAGIC (Warm) and MCBG (Warm): In~\cite{chen2023continuous}, warm-start learning takes an older model and continues training the entire model with new samples. However, we found that freezing the weights of the initial layer of the older model performs better on Big-15, so we opted for this approach. In fact, this approach aligns with a commonly used fine-tuning paradigm in transfer learning, where certain layers are frozen to retain the knowledge gained from the source data.

DAN: The network architecture of the feature extractor and classifier is the same as the generator and classifier mentioned in Appendix~\ref{app_our}.  The feature extractor has $3$ \verb"GIN" layers and ends with a global average pooling, and a dense layer with $256$ neurons. The architecture of the classifier consists of $2$ fully connected layers (\verb"FC_1",  \verb"FC_OUT"). The number of neurons in \verb"FC1" is $256$. \verb"FC_OUT" is the output layer for label prediction. Batch normalization is applied on each hidden layer. For training the model, we use the Adam optimizer with a learning rate of $\mathit{1e-3}$ for $60$ epochs. The batch size is $16$. The coefficient of the MMD loss is set to $1$, followed by~\cite{long2015learning}. We set $\lambda = 0.1$ in $\mathcal{L}_c$ as well.

\subsubsection{Graph-based clustering implementation details}\label{app_graph_clustering}

First, we train a graph autoencoder to derive a 256-dimensional feature vector for each malware graph.  In the consensus clustering algorithm, we apply three different clustering predictors: Gaussian Mixture Model (GMM), HDBSCAN, and K-means. For each predictor, we follow the best practice for selecting the parameters. For K-means clustering, we use a heuristic approach based on inertia values to determine the appropriate cluster number.  In the case of GMM, all combinations of six components and four covariance types are explored to identify the model with the lowest Bayes Information Criterion. Regarding HDBSCAN,  in order to reduce the number of outliers, we set the parameters as follows: $min\_cluster\_size = 2$ and $min\_samples = 2$. All three clustering solutions and the original labels are considered during the consensus matrix update. Finally, the GMM model is utilized to derive the final clusters for our analysis. 

\subsubsection{Additional results}

\begin{figure*}[h]
  \centering
  \includegraphics[width=0.7\linewidth]{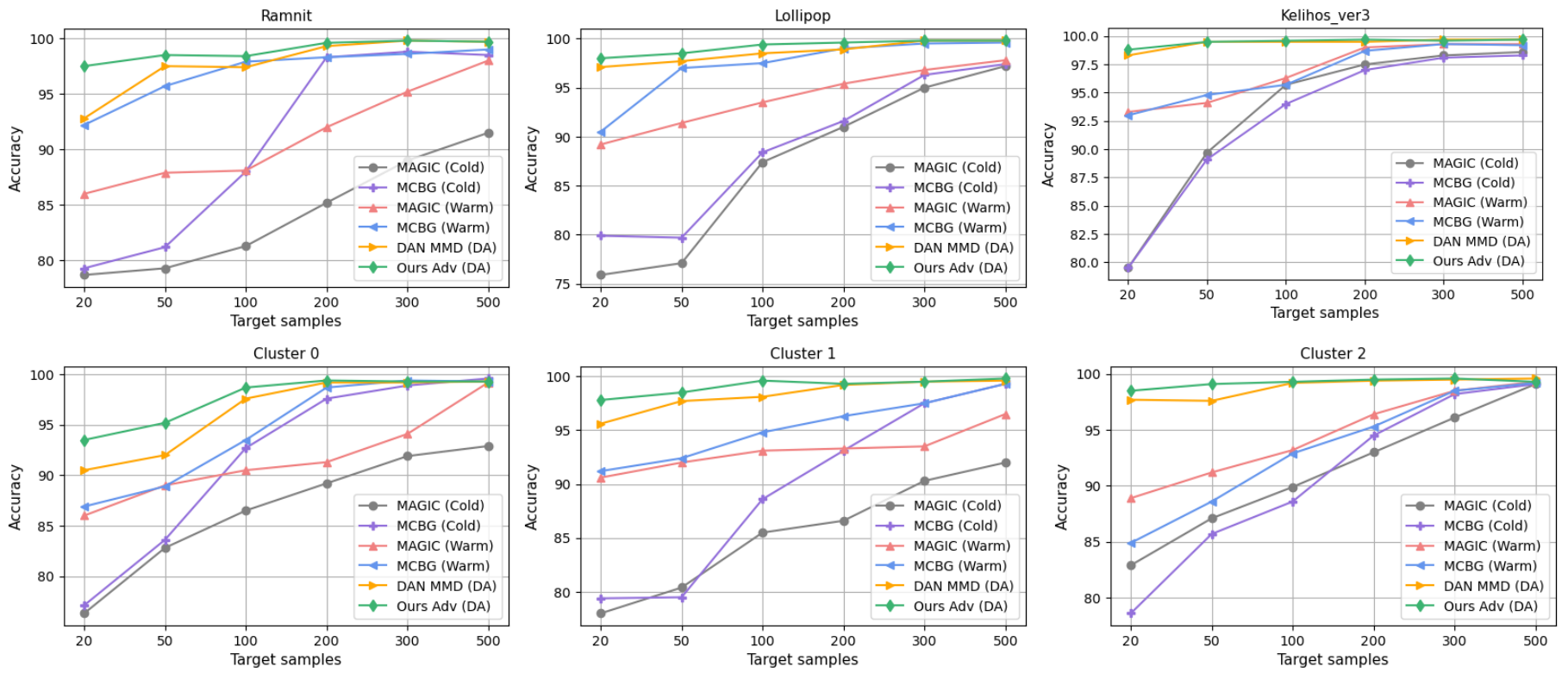}
  \caption{Given fixed target training labels, we compute the average accuracy on the target testing data for different baseline techniques and our method. The top three figures are evaluated on the target family based on the original labels of Big-15, and the bottom three figures are for the clustering labels.}
  \label{app_big15}
\end{figure*}

The F1 scores of all the baselines and our approach evaluated on the original label set of Big-15 are listed in Table~\ref{app_big15_f1}, while the F1 scores for the evaluation on the clustering label set are listed in Table~\ref{app_big15_cluster_f1}. The accuracy of all the methods on each target family based on the original label set and cluster label set of Big-15 is presented in Figure~\ref{app_big15}. The difference between the
averaged accuracy of each approach based on the original label
and cluster label is shown in Table~\ref{avg_acc_diff}.

%\ElisaText{The legend seems to be missing from the figure}

\begin{table}[h]
\caption{Impact of more accurate labels on various training strategies. The table shows the difference in average accuracy, determined by the original and cluster label assignment. A positive value indicates an increase in performance relative to the original label, while a negative value indicates a decrease.}
\label{avg_acc_diff}
\centering
\resizebox{0.8\linewidth}{!}{
\begin{tabular}{@{}ccrrrrrr@{}}
\toprule
\multirow{2}{*}{\textbf{Strategy}} & \multirow{2}{*}{\textbf{Method}} & \multicolumn{6}{c}{\textbf{Target samples}}                                                                                                             \\ \cmidrule(l){3-8} 
                                   &                                  & \multicolumn{1}{l}{20} & \multicolumn{1}{c}{50} & \multicolumn{1}{c}{100} & \multicolumn{1}{c}{200} & \multicolumn{1}{c}{300} & \multicolumn{1}{c}{500} \\ \midrule
\multirow{2}{*}{Cold}              & MCBG                             & -1.2                   & -0.4                   & -1.2                    & -0.5                    & 0.5                     & 1.2                     \\
                                   & MAGIC                            & 1.1                    & 1.4                    & -0.8                    & -1.6                    & -1.3                    & -1.1                    \\ \midrule
\multirow{2}{*}{Warm}              & MCBG                             & -4.2                   & -5.8                   & -3.3                    & -1.9                    & -0.6                    & 0                       \\
                                   & MAGIC                            & -1                     & -0.4                   & -0.3                    & -1.8                    & -1.7                    & -0.1                    \\ \midrule
\multirow{2}{*}{DA}                & DAN (MMD)                        & -1.5                   & -2.4                   & -0.2                    & 0.1                     & -0.4                    & -0.2                    \\
                                   & Ours (Adv)                       & -1.5                   & -1.2                   & 0.1                     & -0.2                    & -0.2                    & -0.2                    \\ \bottomrule
\end{tabular}}
\end{table}

% \usepackage{multirow}
\begin{table*}[t]
\caption{Evaluation results (F1 score) based on the original label set of Big-15.}
\label{app_big15_f1}
\centering
\resizebox{0.8\textwidth}{!}{
\begin{tabular}{@{}ccrrrrrrrrrrrrrrrrrr@{}}
\toprule
                                    &                                   & \multicolumn{6}{c}{\textbf{Ramnit: target samples}}                                                                                                                                                                                                                                                                                                                                                             & \multicolumn{6}{c}{\textbf{Lollipop: target samples}}                                                                                                                                                                                                                                                                                                                                                             & \multicolumn{6}{c}{\textbf{Kelihos\_ver3: target samples}}                                                                                                                                                                                                                                                                                                                                                      \\ \cmidrule(l){3-20} 
\multirow{-2}{*}{\textbf{Strategy}} & \multirow{-2}{*}{\textbf{Method}} & \multicolumn{1}{c}{20}                                            & \multicolumn{1}{c}{50}                                          & \multicolumn{1}{c}{100}                                           & \multicolumn{1}{c}{200}                                           & \multicolumn{1}{c}{300}                                         & \multicolumn{1}{c}{500}                                         & \multicolumn{1}{c}{20}                                            & \multicolumn{1}{c}{50}                                            & \multicolumn{1}{c}{100}                                           & \multicolumn{1}{c}{200}                                           & \multicolumn{1}{c}{300}                                         & \multicolumn{1}{c}{500}                                         & \multicolumn{1}{c}{20}                                            & \multicolumn{1}{c}{50}                                          & \multicolumn{1}{c}{100}                                           & \multicolumn{1}{c}{200}                                           & \multicolumn{1}{c}{300}                                         & \multicolumn{1}{c}{500}                                         \\ \midrule
                                    & MCBG                              & \cellcolor[HTML]{FFFFFF}79.3                                      & \cellcolor[HTML]{FFFFFF}81.3                                    & \cellcolor[HTML]{FFFFFF}88.1                                      & \cellcolor[HTML]{FFFFFF}98.3                                      & \cellcolor[HTML]{FFFFFF}98.2                                    & \cellcolor[HTML]{FFFFFF}98.5                                    & 79.9                                                              & 79.7                                                              & 88.3                                                              & 91.6                                                              & 96.2                                                            & 97.4                                                            & 80.0                                                              & 89.1                                                            & 94.0                                                              & 97.0                                                              & 98.0                                                            & 98.3                                                            \\
\multirow{-2}{*}{Cold}              & MAGIC                             & \cellcolor[HTML]{FFFFFF}78.6                                      & \cellcolor[HTML]{FFFFFF}79.3                                    & \cellcolor[HTML]{FFFFFF}81.2                                      & \cellcolor[HTML]{FFFFFF}85.3                                      & \cellcolor[HTML]{FFFFFF}89.0                                    & \cellcolor[HTML]{FFFFFF}91.5                                    & 75.9                                                              & 77.1                                                              & 87.4                                                              & 91.0                                                              & 95.0                                                            & 97.2                                                            & 79.5                                                              & 89.7                                                            & 95.7                                                              & 97.5                                                              & 98.3                                                            & 98.6                                                            \\ \midrule
                                    & MCBG                              & \cellcolor[HTML]{FFFFFF}92.2                                      & \cellcolor[HTML]{FFFFFF}95.7                                    & \cellcolor[HTML]{FFFFFF}\textbf{97.9}                             & \cellcolor[HTML]{FFFFFF}98.3                                      & \cellcolor[HTML]{FFFFFF}98.5                                    & \cellcolor[HTML]{FFFFFF}99.0                                    & 90.5                                                              & 97.0                                                              & 97.5                                                              & 99.0                                                              & 99.5                                                            & 99.6                                                            & 93.0                                                              & 94.8                                                            & 95.7                                                              & 98.7                                                              & 99.3                                                            & 99.2                                                            \\
\multirow{-2}{*}{Warm}              & MAGIC                             & \cellcolor[HTML]{FFFFFF}86.0                                      & \cellcolor[HTML]{FFFFFF}87.8                                    & \cellcolor[HTML]{FFFFFF}88.2                                      & \cellcolor[HTML]{FFFFFF}92.0                                      & \cellcolor[HTML]{FFFFFF}95.2                                    & \cellcolor[HTML]{FFFFFF}98.0                                    & 89.1                                                              & 91.4                                                              & 93.4                                                              & 95.4                                                              & 96.8                                                            & 97.8                                                            & 93.3                                                              & 94.0                                                            & 96.3                                                              & 99.0                                                              & 99.3                                                            & 99.3                                                            \\ \midrule
                                    & DAN (MMD)                         & \cellcolor[HTML]{FFFFFF}\textbf{92.8}                             & \cellcolor[HTML]{FFFFFF}\textbf{97.5}                           & \cellcolor[HTML]{FFFFFF}97.5                                      & \cellcolor[HTML]{FFFFFF}\textbf{99.3}                             & \cellcolor[HTML]{FFFFFF}\textbf{99.8}                           & \cellcolor[HTML]{FFFFFF}\textbf{99.7}                           & \textbf{97.1}                                                     & \textbf{97.6}                                                     & \textbf{98.5}                                                     & \textbf{98.9}                                                     & \textbf{99.8}                                                   & \textbf{99.8}                                                   & \textbf{98.3}                                                     & \textbf{99.5}                                                            & \textbf{99.5}                                                     & \textbf{99.4}                                                     & \textbf{99.6}                                                   & \textbf{99.7}                                                   \\
                                    &                                   & \cellcolor[HTML]{FFFFFF}\textbf{97.5}                             & \cellcolor[HTML]{FFFFFF}\textbf{98.5}                           & \cellcolor[HTML]{FFFFFF}\textbf{98.4}                             & \cellcolor[HTML]{FFFFFF}\textbf{99.6}                             & \cellcolor[HTML]{FFFFFF}\textbf{99.8}                           & \cellcolor[HTML]{FFFFFF}\textbf{99.7}                           & \textbf{98.0}                                                     & \textbf{98.5}                                                     & \textbf{99.6}                                                     & \textbf{99.6}                                                     & \textbf{99.8}                                                   & \textbf{99.8}                                                   & \textbf{98.5}                                                     & \textbf{99.5}                                                   & \textbf{99.6}                                                     & \textbf{99.7}                                                     & \textbf{99.6}                                                   & \textbf{99.7}                                                   \\
\multirow{-3}{*}{DA}                & \multirow{-2}{*}{Ours (Adv)}      & \multicolumn{1}{l}{{\color[HTML]{32CB00} \textbf{$\uparrow$4.7}}} & \multicolumn{1}{l}{{\color[HTML]{32CB00} \textbf{$\uparrow$1}}} & \multicolumn{1}{l}{{\color[HTML]{32CB00} \textbf{$\uparrow$0.5}}} & \multicolumn{1}{l}{{\color[HTML]{32CB00} \textbf{$\uparrow$0.3}}} & \multicolumn{1}{l}{{\color[HTML]{32CB00} \textbf{$\uparrow$0}}} & \multicolumn{1}{l}{{\color[HTML]{32CB00} \textbf{$\uparrow$0}}} & \multicolumn{1}{l}{{\color[HTML]{32CB00} \textbf{$\uparrow$0.9}}} & \multicolumn{1}{l}{{\color[HTML]{32CB00} \textbf{$\uparrow$0.9}}} & \multicolumn{1}{l}{{\color[HTML]{32CB00} \textbf{$\uparrow$1.1}}} & \multicolumn{1}{l}{{\color[HTML]{32CB00} \textbf{$\uparrow$0.7}}} & \multicolumn{1}{l}{{\color[HTML]{32CB00} \textbf{$\uparrow$0}}} & \multicolumn{1}{l}{{\color[HTML]{32CB00} \textbf{$\uparrow$0}}} & \multicolumn{1}{l}{{\color[HTML]{32CB00} \textbf{$\uparrow$0.2}}} & \multicolumn{1}{l}{{\color[HTML]{32CB00} \textbf{$\uparrow$0}}} & \multicolumn{1}{l}{{\color[HTML]{32CB00} \textbf{$\uparrow$0.1}}} & \multicolumn{1}{l}{{\color[HTML]{32CB00} \textbf{$\uparrow$0.3}}} & \multicolumn{1}{l}{{\color[HTML]{32CB00} \textbf{$\uparrow$0}}} & \multicolumn{1}{l}{{\color[HTML]{32CB00} \textbf{$\uparrow$0}}} \\ \bottomrule
\end{tabular}}
\end{table*}

% \usepackage{multirow}
\begin{table*}[t]
\caption{Evaluation results (F1 score) based on the cluster label of Big-15.}
\label{app_big15_cluster_f1}
\centering
\resizebox{0.8\textwidth}{!}{
\begin{tabular}{@{}ccrrrrrrrrrrrrrrrrrr@{}}
\toprule
                                    &                                   & \multicolumn{6}{c}{\textbf{Cluster 0: target samples}}                                                                                                                                                                                                                                            & \multicolumn{6}{c}{\textbf{Cluster 1: target samples}}                                                                                                                                                                                                                                        & \multicolumn{6}{c}{\textbf{Cluster 2: target samples}}                                                                                                                                                                                                                                          \\ \cmidrule(l){3-20} 
\multirow{-2}{*}{\textbf{Strategy}} & \multirow{-2}{*}{\textbf{Method}} & \multicolumn{1}{c}{20}                        & \multicolumn{1}{c}{50}                        & \multicolumn{1}{c}{100}                       & \multicolumn{1}{c}{200}                       & \multicolumn{1}{c}{300}                         & \multicolumn{1}{c}{500}                         & \multicolumn{1}{c}{20}                        & \multicolumn{1}{c}{50}                          & \multicolumn{1}{c}{100}                       & \multicolumn{1}{c}{200}                       & \multicolumn{1}{c}{300}                     & \multicolumn{1}{c}{500}                       & \multicolumn{1}{c}{20}                        & \multicolumn{1}{c}{50}                        & \multicolumn{1}{c}{100}                       & \multicolumn{1}{c}{200}                       & \multicolumn{1}{c}{300}                       & \multicolumn{1}{c}{500}                         \\ \midrule
                                    & MCBG                              & 77.1                                          & 83.6                                          & 92.7                                          & 97.6                                          & 98.9                                            & 99.6                                            & 79.6                                          & 78.9                                            & 88.5                                          & 93.1                                          & 97.5                                        & 99.3                                          & 78.6                                          & 85.7                                          & 88.6                                          & 94.5                                          & 98.1                                          & 99.2                                            \\
\multirow{-2}{*}{Cold}              & MAGIC                             & 76.3                                          & 82.8                                          & 86.5                                          & 89.2                                          & 91.9                                            & 92.9                                            & 78.0                                          & 80.4                                            & 85.1                                          & 86.6                                          & 90.3                                        & 92.0                                          & 82.9                                          & 87.0                                          & 89.9                                          & 93.0                                          & 96.1                                          & 99.1                                            \\ \midrule
                                    & MCBG                              & 86.9                                          & 88.9                                          & 93.5                                          & 98.6                                          & \textbf{99.4}                                   & \textbf{99.4}                                   & 91.1                                          & 92.4                                            & 94.8                                          & 96.3                                          & 97.5                                        & 99.3                                          & 85.0                                          & 88.6                                          & 93.0                                          & 95.5                                          & 98.5                                          & 99.2                                            \\
\multirow{-2}{*}{Warm}              & MAGIC                             & 86.0                                          & 89.0                                          & 90.5                                          & 91.3                                          & 94.1                                            & 99.2                                            & 90.6                                          & 92.0                                            & 93.1                                          & 93.3                                          & 93.5                                        & 96.5                                          & 88.0                                          & 91.2                                          & 93.2                                          & 96.5                                          & 98.5                                          & 99.3                                            \\ \midrule
                                    & DAN (MMD)                         & \textbf{90.7}                                 & \textbf{92.1}                                 & \textbf{97.6}                                 & \textbf{99.2}                                 & 99.2                                            & 99.3                                            & \textbf{95.7}                                 & \textbf{97.7}                                   & \textbf{98.1}                                 & \textbf{99.2}                                 & \textbf{99.5}                               & \textbf{99.6}                                 & \textbf{97.7}                                 & \textbf{97.6}                                 & \textbf{99.2}                                 & \textbf{99.4}                                 & \textbf{99.5}                                 & \textbf{99.6}                                   \\
                                    &                                   & \textbf{93.4}                                 & \textbf{95.2}                                 & \textbf{98.7}                                 & \textbf{99.4}                                 & \textbf{99.2}                                   & \textbf{99.3}                                   & \textbf{97.8}                                 & \textbf{96.5}                                   & \textbf{99.5}                                 & \textbf{99.3}                                 & \textbf{99.5}                               & \textbf{99.8}                                 & \textbf{98.5}                                 & \textbf{99.1}                                 & \textbf{99.3}                                 & \textbf{99.5}                                 & \textbf{99.6}                                 & \textbf{99.3}                                   \\
\multirow{-3}{*}{DA}                & \multirow{-2}{*}{Ours (Adv)}      & {\color[HTML]{32CB00} \textbf{$\uparrow$2.7}} & {\color[HTML]{32CB00} \textbf{$\uparrow$3.1}} & {\color[HTML]{32CB00} \textbf{$\uparrow$1.1}} & {\color[HTML]{32CB00} \textbf{$\uparrow$0.2}} & {\color[HTML]{CB0000} \textbf{$\downarrow$0.2}} & {\color[HTML]{CB0000} \textbf{$\downarrow$0.1}} & {\color[HTML]{32CB00} \textbf{$\uparrow$2.1}} & {\color[HTML]{CB0000} \textbf{$\downarrow$1.2}} & {\color[HTML]{32CB00} \textbf{$\uparrow$1.4}} & {\color[HTML]{32CB00} \textbf{$\uparrow$0.1}} & {\color[HTML]{32CB00} \textbf{$\uparrow$0}} & {\color[HTML]{32CB00} \textbf{$\uparrow$0.2}} & {\color[HTML]{32CB00} \textbf{$\uparrow$0.8}} & {\color[HTML]{32CB00} \textbf{$\uparrow$1.5}} & {\color[HTML]{32CB00} \textbf{$\uparrow$0.1}} & {\color[HTML]{32CB00} \textbf{$\uparrow$0.1}} & {\color[HTML]{32CB00} \textbf{$\uparrow$0.1}} & {\color[HTML]{CB0000} \textbf{$\downarrow$0.3}} \\ \bottomrule
\end{tabular}
}
\end{table*}

\subsection{Additional details on the content-based features of Big-15}\label{details_content_features}

The categories of features based on the content of the assembly files are summarized below. 

% \begin{table}[h]
% \caption{Features extracted based on the content of assembly file. }
% \label{tabular_feature}
% \centering
% \resizebox{0.8\linewidth}{!}{
% \begin{tabular}{@{}cl@{}}
% \toprule
% Feature Type & \multicolumn{1}{c}{Feature Description}                                                 \\ \midrule
% Symbol       & Frequencies of special symbols                                                          \\
% Opcode       & Frequency of common opcodes                                                             \\
% Register     & Frequency of registers                                                                  \\
% Windows API  & Frequency of use of Windows APIs                                                        \\
% Section      & Statistical attributes that capture the proportion of each section                      \\
% Data Define  & Statistical features that summarize the characteristics of data definition instructions \\
% Others       & File size and the number of lines                                                       \\ \bottomrule
% \end{tabular}}
% \end{table}

\begin{itemize}
  \item Symbol. The frequencies of symbols -, +, *, ], [, ?, and @ are counted due to their common occurrence in code designed to evade detection.
  \item Opcode. We have chosen a subset of $93$ opcodes from the x86 instruction set, selected for their common usage and their frequent appearance in malicious code. We then measure the frequency of these opcodes in each malware sample.
  \item Register. The utilization frequency of registers has proven valuable in classifying malware families. Consequently, we have integrated $26$ register features into the feature set for this purpose.
  \item Windows API. We also measure the frequency of use of Windows API in the assembly code. We have selected the top $794$ frequent APIs used by malicious binaries based on the study in~\cite{ahmadi2016novel}.  
  \item Section. A PE file typically comprises predefined sections such as .text, .data, .bss, .rdata, .edata, .idata, .rsrc, .tls, and .reloc. Due to packing, these default sections can be altered, rearranged, and new sections may be introduced. Consequently, $26$ section features are statistical attributes that capture the proportion of each section within the entire assembly file.
  \item Data Define. Certain malware samples may lack any API calls and primarily consist of a few operation codes, often due to packing. Specifically, they frequently feature data definition instructions like db, dw, and dd. As a result, we have incorporated $18$ statistical features that summarize the distinct characteristics of data definition instructions into our analysis.
  \item Others. We have calculated both the file size and the number of lines within the file.
  
\end{itemize}

\subsection{Implementation details on content-based baselines and our approach}\label{implementation_content}

\subsubsection{SVM and MLP} For SVM, we set $C=0.1$ to be consistent with~\cite{chen2023continuous}. The MLP consists of $8$ fully connected layers (\verb"FC_1",\verb"FC_1",..., \verb"FC_OUT"), with $100$ neurons in \verb"FC1-5" and $400$ neurons in \verb"FC5-6", while \verb"FC_OUT" serves as the output layer for label prediction. MLP is trained using the Adam optimizer with a learning rate of $\mathit{1e-3}$ for $60$ epochs. In the warm-start of MLP, we continue training all the layers of the older model. 

\subsubsection{Our approach} Our model is implemented to be comparable with the MLP. The generator comprises $5$ dense layers, each with $100$ neurons. The classifier architecture consists of $3$ fully connected layers, with the first two layers having $400$ neurons each and the final layer serving as the output layer. Essentially, the combined generator and classifier components form the classifier model described above. The discriminator has four layers with $400$ hidden units, followed by the softmax layer for domain prediction. The hyperparameters remain consistent with those outlined in Appendix~\ref{app_our}.

\subsubsection{DAN} The feature extractor comprises $5$ dense layers, each with $100$ neurons. The classifier architecture consists of $3$ fully connected layers, with the first two layers having $400$ neurons each and the final layer serving as the output layer.  
The hyperparameters remain consistent with those outlined in Appendix~\ref{app_baseline_big15}.

\subsubsection{Additional results on representation evaluation}
The F1-score can be found in Table~\ref{app-representation}.

% \usepackage{multirow}
\begin{table}[t]
\caption{F1 of baselines across various malware representations. In terms of content and image representation, we illustrate the percentage change in accuracy from our adversarial (Adv) DA method to the peak performance among all baselines within the same representation. Ultimately, we demonstrate the enhancements of our full pipeline (graph + Adv DA) compared to both content + Adv DA and image + Adv DA.}
\label{app-representation}
\centering
\resizebox{0.8\linewidth}{!}{
\begin{tabular}{@{}cccrrrrrr@{}}
\toprule
\multicolumn{1}{l}{}                                                                                            &                                             &                                                         & \multicolumn{6}{c}{\textbf{Target samples}}                                                                                                                                                                                                                                                     \\ \cmidrule(l){4-9} 
\multicolumn{1}{l}{\multirow{-2}{*}{\textbf{\begin{tabular}[c]{@{}l@{}}Malware\\ Representation\end{tabular}}}} & \multirow{-2}{*}{\textbf{Strategy}}         & \multirow{-2}{*}{\textbf{Method}}                       & \multicolumn{1}{c}{20}                        & \multicolumn{1}{c}{50}                        & \multicolumn{1}{c}{100}                       & \multicolumn{1}{c}{200}                       & \multicolumn{1}{c}{300}                       & \multicolumn{1}{c}{500}                         \\ \midrule
\multicolumn{1}{c|}{}                                                                                           & \multicolumn{1}{c|}{}                       & \multicolumn{1}{c|}{SVM}                                & 67.1                                          & 71.4                                          & 75.1                                          & 78.9                                          & 82.2                                          & 85                                              \\ \cmidrule(lr){3-3}
\multicolumn{1}{c|}{}                                                                                           & \multicolumn{1}{c|}{\multirow{-2}{*}{Cold}} & \multicolumn{1}{c|}{MLP}                                & 77                                            & 79.4                                          & 85.2                                          & 91.5                                          & 92.9                                          & 94.8                                            \\ \cmidrule(lr){2-3}
\multicolumn{1}{c|}{}                                                                                           & \multicolumn{1}{c|}{}                       & \multicolumn{1}{c|}{SVM}                                & 70.2                                          & 72.7                                          & 78.6                                          & 82.9                                          & 86.7                                          & 89.7                                            \\ \cmidrule(lr){3-3}
\multicolumn{1}{c|}{}                                                                                           & \multicolumn{1}{c|}{\multirow{-2}{*}{Warm}} & \multicolumn{1}{c|}{MLP}                                & \textbf{90.9}                                 & 93.4                                          & 95.3                                          & 96.2                                          & \textbf{97.5}                                 & \textbf{97.9}                                   \\ \cmidrule(lr){2-3}
\multicolumn{1}{c|}{}                                                                                           & \multicolumn{1}{c|}{}                       & \multicolumn{1}{c|}{DAN (MMD) + MLP}                    & 90.8                                          & \textbf{93.9}                                 & \textbf{95.4}                                 & \textbf{96.9}                                 & 97.1                                          & 97.5                                            \\ \cmidrule(lr){3-3}
\multicolumn{1}{c|}{}                                                                                           & \multicolumn{1}{c|}{}                       & \multicolumn{1}{c|}{}                                   & \textbf{94.1}                                 & \textbf{95.2}                                 & \textbf{96.2}                                 & \textbf{97.1}                                 & \textbf{97.7}                                 & \textbf{97.8}                                   \\
\multicolumn{1}{c|}{\multirow{-7}{*}{\begin{tabular}[c]{@{}c@{}}Content-based\\ (CB)\end{tabular}}}             & \multicolumn{1}{c|}{\multirow{-3}{*}{DA}}   & \multicolumn{1}{c|}{\multirow{-2}{*}{Ours (Adv) + MLP}} & {\color[HTML]{32CB00} \textbf{$\uparrow$3.2}} & {\color[HTML]{32CB00} \textbf{$\uparrow$1.3}} & {\color[HTML]{32CB00} \textbf{$\uparrow$0.8}} & {\color[HTML]{32CB00} \textbf{$\uparrow$0.2}} & {\color[HTML]{32CB00} \textbf{$\uparrow$0.2}} & {\color[HTML]{FE0000} \textbf{$\downarrow$0.1}} \\ \midrule
\multicolumn{1}{c|}{}                                                                                           & \multicolumn{1}{c|}{Cold}                   & \multicolumn{1}{c|}{ResNet-50}                          & 60.6                                          & 60.6                                          & 70.6                                          & 74.3                                          & 77                                            & 84.4                                            \\ \cmidrule(lr){2-3}
\multicolumn{1}{c|}{}                                                                                           & \multicolumn{1}{c|}{Warm}                   & \multicolumn{1}{c|}{ResNet-50}                          & \textbf{86.2}                                 & \textbf{87.7}                                 & 89.5                                          & \textbf{91.9}                                 & \textbf{93.5}                                 & \textbf{94.4}                                   \\ \cmidrule(lr){2-3}
\multicolumn{1}{c|}{}                                                                                           & \multicolumn{1}{c|}{}                       & \multicolumn{1}{c|}{DAN (MMD) + CNN}                    & 85.4                                          & 87.5                                          & \textbf{89.9}                                 & 91.6                                          & 93.1                                          & 94                                              \\ \cmidrule(lr){3-3}
\multicolumn{1}{c|}{}                                                                                           & \multicolumn{1}{c|}{}                       & \multicolumn{1}{c|}{}                                   & \textbf{88.6}                                 & \textbf{90}                                   & \textbf{92.2}                                 & \textbf{93.1}                                 & \textbf{94}                                   & \textbf{94.6}                                   \\
\multicolumn{1}{c|}{\multirow{-5}{*}{\begin{tabular}[c]{@{}c@{}}Image-based\\ (IB)\end{tabular}}}               & \multicolumn{1}{c|}{\multirow{-3}{*}{DA}}   & \multicolumn{1}{c|}{\multirow{-2}{*}{Ours (Adv) + CNN}} & {\color[HTML]{32CB00} \textbf{$\uparrow$2.4}} & {\color[HTML]{32CB00} \textbf{$\uparrow$2.3}} & {\color[HTML]{32CB00} \textbf{$\uparrow$2.3}} & {\color[HTML]{32CB00} \textbf{$\uparrow$1.2}} & {\color[HTML]{32CB00} \textbf{$\uparrow$0.5}} & {\color[HTML]{32CB00} \textbf{$\uparrow$0.2}}   \\ \midrule
\multicolumn{1}{c|}{\begin{tabular}[c]{@{}c@{}}Graph-based\\ (GB)\end{tabular}}                                 & \multicolumn{1}{c|}{DA}                     & \multicolumn{1}{c|}{Ours (Adv) + GIN}                   & \textbf{96.6}                                 & \textbf{96.9}                                 & \textbf{99.2}                                 & \textbf{99.4}                                 & \textbf{99.4}                                 & \textbf{99.5}                                   \\ \midrule
\multicolumn{3}{c}{\textbf{Improvement over CB: Ours (Adv)}}                                                                                                                                                            & {\color[HTML]{32CB00} \textbf{$\uparrow$2.5}} & {\color[HTML]{32CB00} \textbf{$\uparrow$1.7}} & {\color[HTML]{32CB00} \textbf{$\uparrow$3}}   & {\color[HTML]{32CB00} \textbf{$\uparrow$2.3}} & {\color[HTML]{32CB00} \textbf{$\uparrow$1.7}} & {\color[HTML]{32CB00} \textbf{$\uparrow$1.7}}   \\
\multicolumn{3}{c}{\textbf{Improvement over IB: Ours (Adv)}}                                                                                                                                                            & {\color[HTML]{32CB00} \textbf{$\uparrow$8}}   & {\color[HTML]{32CB00} \textbf{$\uparrow$6.9}} & {\color[HTML]{32CB00} \textbf{$\uparrow$7}}   & {\color[HTML]{32CB00} \textbf{$\uparrow$6.3}} & {\color[HTML]{32CB00} \textbf{$\uparrow$5.4}} & {\color[HTML]{32CB00} \textbf{$\uparrow$4.9}}   \\ \bottomrule
\end{tabular}}
\end{table}

\subsection{Implementation details on image-based baselines}
\label{implementation_image}
\subsubsection{Our approach} The generator has the architecture of (\verb"Conv2D", \verb"MaxPooling", \verb"Conv2D", \verb"MaxPooling", \verb"Flatten"). The \verb"Conv2D" layers has $\{32,64\}$ $3 \times 3$ filters respectively.  The classifier consists of $2$ fully connected layers (\verb"FC_1",  \verb"FC_OUT"). The number of neurons in \verb"FC1" is $256$. \verb"FC_OUT" is the output layer for label prediction. The discriminator has two layers with $1024$ hidden units and is followed by the softmax layer for domain prediction. Batch normalization is applied on each hidden layer. For training the model, we use the Adam optimizer with a learning rate of $\mathit{1e-3}$ for $60$ epochs. The batch size is $32$. The coefficient of the loss $\mathcal{L}_g$ is set to $0.1$. We set $\lambda = 0.1$ in $\mathcal{L}_c$.

\subsubsection{ResNet (Cold) and (Warm)} The original ResNet-50's output layer contains $1000$ neurons. As our task focuses on predicting whether a sample is malware or benign software, we modified the model by removing its last layer, adding a global average pooling layer, incorporating a fully connected layer with $256$ neurons, and appending an output layer with $2$ neurons.   For training the model, we use the Adam optimizer with a learning rate of $\mathit{1e-3}$ for $60$ epochs.

\subsubsection{DAN} The feature extractor also has the same architecture as the generator and so as the classifier.  For training the model, we use the Adam optimizer with a learning rate of $\mathit{1e-3}$ for $60$ epochs. The batch size is $32$. The coefficient of the MMD loss is set to $1$. We set $\lambda = 0.1$ in $\mathcal{L}_c$.

\subsubsection{Additional results on representation evaluation}
The F1-scores can be found in Table~\ref{app-representation}.

\subsection{Implementation details and additional results on MB-24}\label{implementation_drift}

\subsubsection{Implementation}\label{app_mb24_imple} The implementation of the baseline models and our approach remain consistent with those outlined in Appendix~\ref{app_our} and \ref{app_baseline_big15}. We set $\lambda = 0.5$ in $\mathcal{L}_c$ for DAN and our approach.

\subsubsection{Additional results on MB-24 evaluation}
The F1-scores can be found in Table~\ref{app-mb_24_result}.

% \usepackage{multirow}
\begin{table}[h]
\caption{Evaluation results (F1 score) on the MB-24 dataset}
\label{app-mb_24_result}
\centering
\resizebox{0.8\linewidth}{!}{
\begin{tabular}{@{}ccrrrrrr@{}}
\toprule
                                    &                                   & \multicolumn{6}{c}{\textbf{Target samples}}                                                                                                                                                                                                                                                 \\ \cmidrule(l){3-8} 
\multirow{-2}{*}{\textbf{Strategy}} & \multirow{-2}{*}{\textbf{Method}} & \multicolumn{1}{c}{20}                        & \multicolumn{1}{c}{50}                      & \multicolumn{1}{c}{100}                       & \multicolumn{1}{c}{200}                       & \multicolumn{1}{c}{300}                     & \multicolumn{1}{c}{500}                         \\ \midrule
                                    & MCBG                              & 53.8                                          & 56.9                                        & 59.2                                          & 62.7                                          & 70.4                                        & 71.6                                            \\
\multirow{-2}{*}{Cold}              & MAGIC                             & 57.4                                          & 61.5                                        & 67.4                                          & 69.2                                          & 70.2                                        & 71.7                                            \\ \midrule
                                    & MCBG                              & 68.1                                          & \textbf{71.7}                               & 73.1                                          & 74.8                                          & 76.7                                        & 77.5                                            \\
\multirow{-2}{*}{Warm}              & MAGIC                             & 68.7                                          & 71.5                                        & 73.4                                          & 75                                            & 76.1                                        & 77.3                                            \\ \midrule
                                    & DAN (MMD)                         & \textbf{70.5}                                 & 71.5                                        & \textbf{73.7}                                 & \textbf{78.4}                                 & \textbf{78.8}                               & \textbf{80.5}                                   \\
                                    &                                   & \textbf{72.1}                                 & \textbf{73.7}                               & \textbf{76.1}                                 & \textbf{79.2}                                 & \textbf{79.8}                               & \textbf{82.4}                                   \\
\multirow{-3}{*}{DA}                & \multirow{-2}{*}{Ours (Adv)}      & {\color[HTML]{32CB00} \textbf{$\uparrow$1.6}} & {\color[HTML]{32CB00} \textbf{$\uparrow$2}} & {\color[HTML]{32CB00} \textbf{$\uparrow$2.4}} & {\color[HTML]{32CB00} \textbf{$\uparrow$0.8}} & {\color[HTML]{32CB00} \textbf{$\uparrow$1}} & {\color[HTML]{32CB00} \textbf{$\uparrow$1.9}} \\ \bottomrule
\end{tabular}}
\end{table}

\subsubsection{Additional results on the obfuscated malware} 

\begin{figure}[t]
  \centering
  \includegraphics[width=\linewidth]{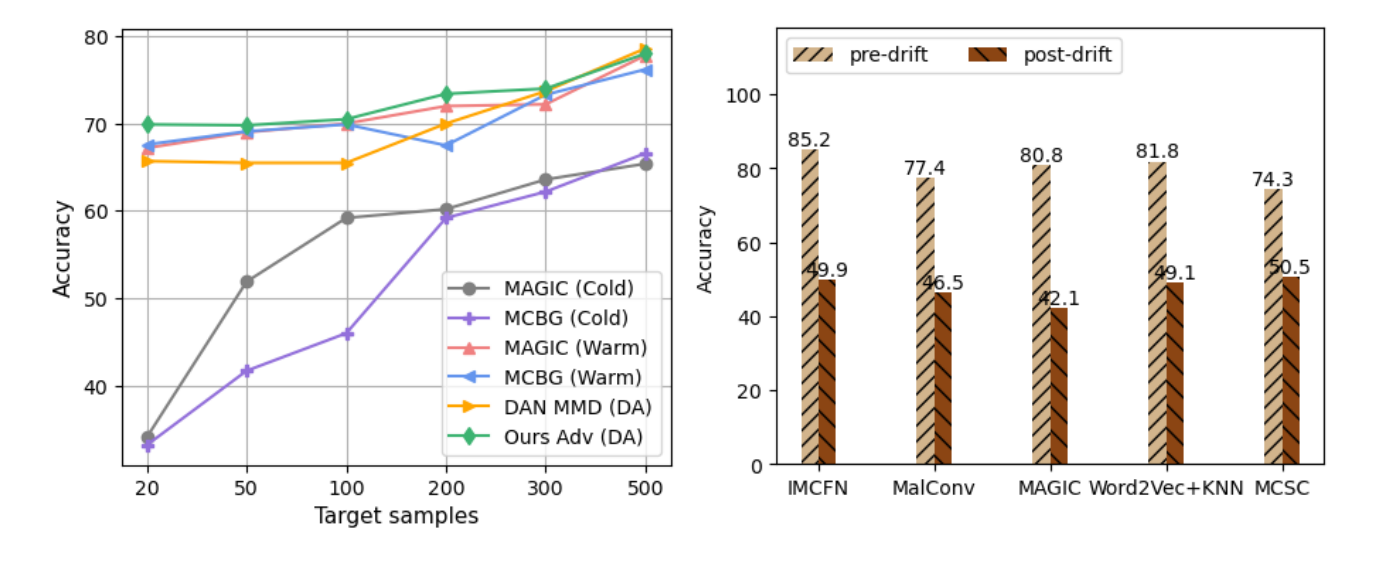}
  \caption{\textbf{Left}: Accuracy on the obfuscated September testing data for different baseline techniques and our method. \textbf{Right}: Accuracy of source-only approach on the MalwareDrift dataset.}
  \label{app_mb24_obf_ma}
\end{figure}

The first experiment was conducted using task (b) in Figure~\ref{mb24}. The source malware data was collected from March to May, the target training malware was from August, and the target testing malware was from September. All target testing malware samples were obfuscated using Hyperion, while the source and target training malware samples remained unobfuscated. Neither the baselines nor our model were exposed to obfuscated samples during training. The goal was to compare the performance of each method, trained with unobfuscated malware when tested on obfuscated malware. The results of this experiment are reported in Figure~\ref{app_mb24_obf_ma}. The experiment shows that our approach experiences the least accuracy degradation when tested on obfuscated malware samples that were not seen during training. For a comparison, please refer to Figure~\ref{mb24_result}, which reports the results when the target testing set is not obfuscated.

% \begin{figure}[t]
%   \centering
%   \includegraphics[width=0.6\linewidth]{figures/mb-24-obf-app.pdf}
%   \caption{Accuracy on the obfuscated September testing data for different baseline techniques and our method.  }
%   \label{app_mb24_obf}
% \end{figure}

% \begin{figure}[h]
%   \centering
%   \includegraphics[width=0.6\linewidth]{figures/ndss_revision_ma.pdf}
%   \caption{Accuracy of source only approach on the MalwareDrift dataset.}
%   \label{app_malwaredrift}
% \end{figure}

\subsection{Implementation details and additional results on MalwareDrift}\label{implementation_drift}

The implementation of the baseline models and our approach remain consistent with those outlined in Appendix~\ref{app_our} and \ref{app_baseline_big15}, except for the output layer, which is modified to accommodate $8$ classes.

Previous work by Ma et al.~\cite{ma2021comprehensive} has shown that models trained on the pre-drift data perform poorly on the post-drift data.  Figure~\ref{app_mb24_obf_ma}  shows performance data from ~\cite{ma2021comprehensive} on pre-drift and post-drift testing data for five models trained solely on pre-drift training data. As we can see from the figure, the reduction in accuracy ranges from $23.0\%$ to $38.7\%$. It is thus clear that all considered models experience a substantial performance decrease when confronted with code evolution within the same malware families.
%\ElisaText{Sometimes, you use F1 and other times you use F1. I think that we should use F1 and be consistent.}

We report the accuracy and F1 score of all the methods evaluated on post-drift data using $10-45$ labeled data from each post-drift family in Table~\ref{app_multi}.

% \begin{figure}[h]
%   \centering
%   \includegraphics[width=0.6\linewidth]{figures/ndss_revision_ma.pdf}
%   \caption{Accuracy of source only approach on the MalwareDrift dataset.}
%   \label{app_malwaredrift}
% \end{figure}

% \usepackage{multirow}
\begin{table}[h]
\caption{Accuracy and F1 score on post-drift testing data using a few labeled data from each family in post-drift training data.}
\centering
\label{app_multi}
\centering
\resizebox{0.8\linewidth}{!}{
\begin{tabular}{@{}cccrrrrr@{}}
\toprule
                                  &                                     &                                    & \multicolumn{5}{c}{\textbf{Target samples per family}}                                                                                                                                                                                  \\ \cmidrule(l){4-8} 
\multirow{-2}{*}{\textbf{Metric}} & \multirow{-2}{*}{\textbf{Strategy}} & \multirow{-2}{*}{\textbf{Method}}  & \multicolumn{1}{c}{10}                      & \multicolumn{1}{c}{20}                      & \multicolumn{1}{c}{30}                        & \multicolumn{1}{c}{40}                      & \multicolumn{1}{c}{45}                        \\ \midrule
                                  &                                     & MCBG                               & 25.5                                        & 35.3                                        & 40.0                                          & 63.5                                        & 73.2                                          \\
                                  & \multirow{-2}{*}{Cold}              & MAGIC                              & 21.0                                        & 34.5                                        & 39.4                                          & 59.0                                        & 69.5                                          \\
                                  &                                     & MCBG                               & 71                                          & 77.0                                        & 83.0                                          & 85.7                                        & 88.9                                          \\
                                  & \multirow{-2}{*}{Warm}              & MAGIC                              & 69.0                                        & 74.5                                        & 82.0                                          & 83.5                                        & 83.5                                          \\
                                  &                                     & DAN (MMD) + GIN                    & \textbf{74.0}                               & \textbf{84.5}                               & \textbf{86.2}                                 & \textbf{88.0}                               & \textbf{89.0}                                 \\
                                  &                                     &                                    & \textbf{83.0}                               & \textbf{87.5}                               & \textbf{88.7}                                 & \textbf{90.0}                               & \textbf{91.6}                                 \\
\multirow{-7}{*}{Accuracy}        & \multirow{-3}{*}{DA}                & \multirow{-2}{*}{Ours (Adv) + GIN} & {\color[HTML]{32CB00} \textbf{$\uparrow$9}} & {\color[HTML]{32CB00} \textbf{$\uparrow$3}} & {\color[HTML]{32CB00} \textbf{$\uparrow$2.5}} & {\color[HTML]{32CB00} \textbf{$\uparrow$2}} & {\color[HTML]{32CB00} \textbf{$\uparrow$2.6}} \\ \midrule
                                  &                                     & MCBG                               & 25.5                                        & 35.2                                        & 40.0                                          & 63.5                                        & 73.2                                          \\
                                  & \multirow{-2}{*}{Cold}              & MAGIC                              & 22.0                                        & 34.5                                        & 39.4                                          & 58.0                                        & 68.5                               \\
                                  &                                     & MCBG                               & 71.5                                        & 77.0                                        & 82.5                                          & 85.5                                        & 88.5                                          \\
                                  & \multirow{-2}{*}{Warm}              & MAGIC                              & 69.0                                        & 74.5                                        & 81.0                                          & 83.5                                        & 83.5                                          \\
                                  &                                     & DAN (MMD) + GIN                    & \textbf{75.0}                               & \textbf{84.5}                               & \textbf{86.2}                                 & \textbf{88.0}                               & \textbf{89.0}                                 \\
                                  &                                     &                                    & \textbf{82.0}                               & \textbf{87.5}                               & \textbf{88.7}                                 & \textbf{89.0}                               & \textbf{90.7}                                 \\
\multirow{-7}{*}{F1}              & \multirow{-3}{*}{DA}                & \multirow{-2}{*}{Ours (Adv) + GIN} & {\color[HTML]{32CB00} \textbf{$\uparrow$7}} & {\color[HTML]{32CB00} \textbf{$\uparrow$3}} & {\color[HTML]{32CB00} \textbf{$\uparrow$2.5}} & {\color[HTML]{32CB00} \textbf{$\uparrow$1}} & {\color[HTML]{32CB00} \textbf{$\uparrow$1.7}} \\ \bottomrule
\end{tabular}
}
\end{table}

\subsection{Computational runtime}\label{app_runtime}
Both CFG extraction (Section~\ref{cfg}) and vertex feature extraction (Section~\ref{vertex}) are dependent on the node count within the CFG. Therefore, we present the extraction times for a malware CFG containing $17,564$ basic blocks/nodes. The CFG extraction process takes only $2.84$ seconds to extract a complete CFG with $17,564$ basic blocks. In contrast, the vertex feature extraction is the most time-consuming step. For a CFG with $17,564$ nodes, the encoding process using PalmTree requires $3.4$ minutes.

Training, on the other hand, is relatively faster. The training duration is affected by the size of the source and target datasets, the total number of epochs, and the GPU used. For example, training on the MB-24 dataset, which includes $7,907$ source training samples and $200$ target training samples over $80$ epochs, takes $13.5$ minutes on an NVIDIA RTX 4090D. This is highly efficient for training a deep learning model.

\section{Artifact Appendix}

\subsection{Description \& Requirements}

\subsubsection{How to access}
The code for the artifact can be accessed at: \url{https://github.com/gloryer/malware-detection-concept-drift/tree/main?tab=readme-ov-file}. The required data to run this artifact can be accessed at: \url{https://zenodo.org/records/14213306}.

\subsubsection{Hardware dependencies}
We have successfully run the code with the following hardware
\begin{itemize}
    \item CPU: Intel® Core™ i7-6700K CPU @ 4.00GHz × $8$
    \item GPU: Nvidia RTX 3090 ($24$ GB)
    \item Memory: $64$ GB
\end{itemize}

Additionally, we recommend $100$ GB of available disk space to store the data.
\subsubsection{Software dependencies} 
The code was tested on Ubuntu 20.04 LTS and using Python 3.8. It should also run on Ubuntu 22.04 LTS and later stable versions. Most of the code is built with TensorFlow, but PalmTree was developed using PyTorch, so some notebooks also require PyTorch installation. The versions used are TensorFlow 2.9.0 and PyTorch 2.4. Please follow the README in our GitHub repository to set up the environment and solve software dependencies.

\subsection{Artifact Installation \& Configuration}

To set up the environment, we provided a \texttt{requirements.txt} (with the exact versions we used) in our repository. We recommend creating a virtual environment and installing the packages. All the required libraries can be installed with

\texttt{pip install -r requirements.txt}

The requirements were generated using \texttt{pip freeze} and modified manually to consider only the required packages. If a package is missing or you run into a problem, please feel free to contact us.

\subsection{Experiment Workflow}

Once the GitHub repository is cloned, please download the data and store them under the directory \texttt{malware-detection-concept-drift}. Run \texttt{tar -xzvf data.tar.gz} to extract the compressed file and do not change the name of the extracted folder (the name should be \texttt{/data}. Detailed instructions can be found in the GitHub README file. The experiments are designed to help the readers work through our pipeline presented in the paper (see the following evaluation section). We are seeking the artifact available badge and the artifact functional badge for this artifact. To make the review process convenient, we provided ipynb notebooks to run the experiments.

\subsection{Major Claims}
Following are the major claims we make for the artifact available and artifact functional badge

\begin{itemize}
    \item (C1): Our artifact extracts CFGs from assembly files, and the raw CFGs are further processed to generate the node-level embeddings. This is proved by E1 and E2. 
    \item (C2): We find that the original labels of Big-15 are not well separated. We can create denser clusters for the same dataset using the graph-based clustering algorithm presented in Section~\ref{gca}. This is proved by E3 and E4.
    \item (C3): Our artifact can be successfully run to train our domain adaptation models and generate evaluation metrics (to produce our results in Figure~\ref{big15} and Table~\ref{Big-15-representation}) in the paper. This is proved by E5. 
\end{itemize}

\subsection{Evaluation}
Please run the following experiments to verify the claims.

\subsubsection{Experiment (E1)}\label{e1}
[CFG Extraction] [$10$ human-minutes + $5$ compute-minutes]: The following experiment is to extract CFGs from assembly files of malware/benign binaries.

\begin{itemize}
    \item From the \texttt{CFG} directory, run the notebook \texttt{extract\_cfg.ipynb}
    \item We only include five example ASM files from the Big-15 dataset to show the code is functional. However, the code is scalable to process a large number of files. You can also view the node(s) and edges printed in pretty json in the second cell.
\end{itemize}

\subsubsection{Experiment (E2)}\label{e2}
[Vertex Feature Extraction] [$15$ human-minutes + $20$ compute-minutes]: The following experiment is to generate the embeddings for the nodes of CFG using the pre-trained PalmTree model.

\begin{itemize}
    \item From the  \texttt{CFG} directory, run the notebook \texttt{generate\_embeddings.ipynb}
    \item We only include five CFGs to show the code is functional. However, the code is scalable and can perform over a very large number of files.
\end{itemize}

\subsubsection{Experiment (E3)}\label{e3}
[Graph-based Clustering Phase 1] [$20$ human-minutes + $20$ compute-minutes]: The following experiment is to generate the graph embedding with a graph autoencoder. The graph embeddings will be used in Phase 2 of the graph-based clustering.

\begin{itemize}
    \item From the \texttt{graph\_based\_clustering} directory, \\ run the notebook \\ \texttt{generate\_graph\_embeddings.ipynb}
    \item To simplify the amount of work for testing, we demonstrate with just five graphs obtained from E2, but the code is designed to process many graphs.
\end{itemize}

\subsubsection{Experiment (E4)}\label{e4}
[Graph-based Clustering Phase 2] [$1$ human-hour + $1$ compute-hour]: The following experiment is to generate the cluster labels with our weighted consensus clustering algorithm.

\begin{itemize}
    \item From the \texttt{graph\_based\_clustering} directory,\\
    run the notebook \\
    \texttt{consensus\_clustering.ipynb}
    \item Within the notebook, you'll find cells for t-SNE visualizations of the Big-15 dataset, showing both the original labels and the new cluster labels. Each cell will generate a figure in the output, which can be compared to Figure~\ref{tsne}.
\end{itemize}

\subsubsection{Experiment (E5)}\label{e5}
[Domain Adaptation] [$2$ human-hours + $4$ compute-hours]: The following experiment is to generate our models' performance metrics on the Big-15 dataset. Follow these instructions to train our model:
\begin{itemize}
    \item Run the following notebooks in any order
        \texttt{Train\_origin\_label\_graph.ipynb}
        \texttt{Train\_cluster\_label\_graph.ipynb}
        \texttt{Train\_cluster\_label\_image.ipynb}
         \texttt{Train\_cluster\_label\_content.ipynb}
    \item The first two notebooks will generate the metrics for our approach, as shown in Figure~\ref{app_big15}. The third and fourth notebooks will generate the metrics for our approach using image representations and content-based representations, respectively, corresponding to Table~\ref{Big-15-representation}. 
\end{itemize}

\subsection{Notes} Each notebook can be run without any modification to the cells. Since there are many cells, we recommend using Kernel → Run all cells (as found in the Jupyter interface). For quicker and more convenient execution, we set the number of epochs to a low value, minimizing runtime. However, the notebooks can be adjusted to use a larger number of epochs. To do this, follow the cells in the training notebooks, where we've indicated in the code comments how to modify the epoch setting.

\end{document}